\documentclass[11pt,prd,preprint,superscriptaddress]{revtex4}
 \pdfoutput=1
\usepackage{revsymb}
\usepackage{graphicx}
\usepackage{bm}
\usepackage{amsfonts}
\usepackage{amsmath}
\usepackage{amssymb}
\usepackage{graphicx}

\begin{document}

\title{Matter growth in extended $\Lambda$CDM cosmology}

\date{\today}

\author{W. Zimdahl\footnote{E-mail: winfried.zimdahl@pq.cnpq.br}}
\affiliation{Universidade Federal do Esp\'{\i}rito Santo,
Departamento
de F\'{\i}sica\\
Av. Fernando Ferrari, 514, Campus de Goiabeiras, CEP 29075-910,
Vit\'oria, Esp\'{\i}rito Santo, Brazil}
\author{H.E.S. Velten\footnote{E-mail: velten@pq.cnpq.br}}
\affiliation{Universidade Federal do Esp\'{\i}rito Santo,
Departamento
de F\'{\i}sica\\
Av. Fernando Ferrari, 514, Campus de Goiabeiras, CEP 29075-910,
Vit\'oria, Esp\'{\i}rito Santo, Brazil}
\author{W.C. Algoner\footnote{E-mail:  w.algoner@cosmo-ufes.org}}
\affiliation{Universidade Federal do Esp\'{\i}rito Santo,
Departamento
de F\'{\i}sica\\
Av. Fernando Ferrari, 514, Campus de Goiabeiras, CEP 29075-910,
Vit\'oria, Esp\'{\i}rito Santo, Brazil}

\begin{abstract}
On the basis of a previously established scalar-tensor extension of the $\Lambda$CDM model we develop an effective fluid approach for the matter growth function.
This extended $\Lambda$CDM (henceforth $e_{\Phi}\Lambda$CDM) cosmology takes into account deviations from the
standard model both via a modified background expansion and by the inclusion of geometric anisotropic stresses as well as of perturbations of the geometric dark-energy equivalent.
The background dynamics is governed by an explicit analytic expression for the Hubble rate in which modifications of the standard model are given in terms of a single constant parameter \cite{WHW}.
To close the system of fluid-dynamical perturbation equations we introduce two phenomenological parameters through which the anisotropic stress is related both to the total energy density perturbation of the cosmic substratum and to relative perturbations in the effective two-component system.
We quantify the impact of deviations from the standard background, of anisotropic stresses and of non-vanishing perturbations of the effective dark-energy component
on the matter growth rate function $f \sigma_8$ and confront the results with recent redshift-space distortion (RSD) measurements.
\end{abstract}

%\pacs{98.80.-k, 04.50.+h}

\maketitle
\date{\today}

\section{Introduction}
\label{Introduction}

In scalar-tensor theories
the gravitational interaction is mediated both by a
metric tensor and a scalar field.
The prototype of a scalar-tensor theory is
Jordan-Brans-Dicke (JBD) theory \cite{jordan,BD,B}, based on earlier ideas of Mach.
Theories of this type represent generalizations of Einstein's General Relativity (GR).
Studies of different aspects of scalar-tensor theories have been performed, e.g., in
\cite{nagata,catena,farao,dolgov,chiba,duma,sofarao,abean,scalten,clifton,HwangNoh2005,battye,battye13,battye16,chibayam,joyce}.

Scalar-tensor theories aim at explaining the observed late-time accelerated expansion of the scale factor of the Robertson-Walker (RW) metric without the introduction of a dark-energy (DE) component in the matter sector of GR.
Instead, it is the geometric part of the (extended)  field equations which is supposed to provide a geometric DE equivalent
(see, e.g., \cite{torres,carroll,nojiri1,gannouji1,copeland,lobo,caldkam,baker,limaferreira,kofinaslima}).
Limits on the parameters of scalar-tensor theories from cosmological  observations have been obtained, e.g., in \cite{wuchen,liwuchen,hrycyna,avilezskordis,umilta,alonsobellini}.

For many applications it is useful to map the additional (compared with GR) geometrical terms in the scalar-tensor gravitational field equations  onto an effective fluid component.
Formally, the cosmological dynamics is then modeled as an effective two-component system within GR, where one of the components is of geometric origin (see, e.g., \cite{HwangNoh2005,battye,battye13,battye16,baker}).

In a previous paper we established on this basis a scalar-tensor extension of the background $\Lambda$CDM model with an explicit analytic expression for
the Hubble rate \cite{WHW}.
Deviations from the $\Lambda$CDM model are described by a single parameter which also governs the effective scalar-field dynamics.
It is a characteristic feature of this $e_{\Phi}\Lambda$CDM approach that it does not rely on an explicit solution of the underlying scalar-field dynamics.
Instead, the dynamics is solved on the level of conservation equations for the effective fluid energy densities, including the energy density of the geometric ``fluid".
This exploits the circumstance that the scalar field enters the gravitational dynamics only through the effective energy-momentum tensor of this ``fluid" \cite{LAmend}. All the details of the exact scalar-field dynamics are not required here, in particular, no assumptions about the scalar-field potential are necessary.

Our main interest in this paper is the matter growth function which is considered to be a useful tool to discriminate between different theories of gravitation.
In establishing the $e_{\Phi}\Lambda$CDM perturbation analysis we benefit from the existence of an analytically known background dynamics
which determines the coefficients in the perturbation equations.
The relevant equations follow from the separate (in the Jordan frame) energy-momentum conservations of the matter component and of the geometric ``fluid".
Equivalently, the latter may be replaced by the total energy-momentum conservation equations of the cosmic substratum.
As in any two-component dynamics there appear non-adiabatic (isotropic) pressure perturbations which couple the energy-density perturbations of the components (or of the total density) to relative perturbations within the system, even if the components are adiabatic by themselves. Moreover, the geometric ``fluid" of a scalar-tensor theory generally has anisotropic stresses the existence of which makes it differ from the perfect-fluid based GR.

In the spirit of our JBD inspired $e_{\Phi}\Lambda$CDM fluid approach we do not use the general expression for the anisotropic pressure in terms of the  scalar-field perturbation. We close the system of fluid perturbation equations by two phenomenological coefficients which relate the anisotropic stress both to the total density perturbations and to the relative energy density perturbations in the system. These coefficients also parametrize the difference in the gravitational potentials.

In this paper we take into account perturbations of the geometric DE equivalent in a simplified manner, assuming just a simple proportionality relation to the matter perturbations. This provides us with a shortcut to the $e_{\Phi}\Lambda$CDM matter perturbation dynamics without exactly solving the entire coupled system of perturbation equations. Already this rough approximation allows us  to quantify the impact of DE perturbations on the growth function.
The full perturbation dynamics will be the subject of a subsequent paper.

Our aim is to study the combined effect of deviations from the $\Lambda$CDM background dynamics, of effective anisotropic pressures of geometric origin and of non-vanishing (geometric) DE perturbations on the matter growth function. To the best of our knowledge such combined analysis has not been performed before, at least not semi-analytically.

We shall contrast the theoretical $e_{\Phi}\Lambda$CDM growth function $f\sigma_{8}$ against the 18 data points of the `Gold' growth set listed in \cite{nesseris17}. In particular, we set limits on the parameters that quantify the anisotropic stress  and the DE perturbations and reveal a certain degeneracy between these parameters.

The structure of the paper is as follows.
Section~\ref{scalartensor} recalls basic relations of scalar-tensor theory and introduces an effective fluid description of its dynamics. In section~\ref{background} we present the results for the background dynamics, in particular, the explicit analytic expression for the Hubble rate of the scalar-tensor extension of the standard model \cite{WHW}. Essential relations of the general perturbation dynamics are briefly reviewed  in section~\ref{genpert}.
Section~\ref{aipressure} is devoted to the description of the anisotropic pressure. It introduces the phenomenological parameters through which the perturbation dynamics is closed.
On this basis the matter perturbations are calculated in section~\ref{matter}. Taking into account DE perturbations
in a simplified manner, we confront the results of our fluid-dynamical $e_{\Phi}\Lambda$CDM approach with those of the standard model and with $f\sigma_{8}$ data in section~\ref{data}.
Section~\ref{discussion} summarizes and discusses our results.

\section{JBD inspired effective fluid description}
\label{scalartensor}
\subsection{Effective GR-type description}
Starting from a Jordan-Brans-Dicke type action (see, e.g., \cite{abean,clifton,chibayam})
\begin{equation}
\label{action1}
   S(g_{\mu\nu},\Phi)= \frac{1}{2\kappa^{2}}\int d^{4}x\sqrt{-g}\left[\Phi R - \frac{\omega(\Phi)}{\Phi}\left(\nabla \Phi\right)^{2} - U(\Phi)\right] + S_{m}\left(g_{\mu\nu}\right),
\end{equation}
the Jordan-frame gravitational field equations for scalar-tensor theories are
\begin{eqnarray}
\label{eqJ}
\Phi\left(R_{\mu\nu}- \frac{1}{2}g_{\mu\nu}R\right) &=& \kappa^{2}T_{(m)\mu\nu}
\nonumber\\
&& + \frac{\omega(\Phi)}{\Phi}
\left(\partial_{\mu}\Phi\partial_{\nu}\Phi
- \frac{1}{2}g_{\mu\nu}\left(\nabla\Phi\right)^{2}\right)
+ \nabla_{\mu}\nabla_{\nu}\Phi- g_{\mu\nu}\Box\Phi
- \frac{1}{2}g_{\mu\nu}U,
\end{eqnarray}
and
\begin{equation}\label{boxphi}
\Box \Phi = \frac{1}{2\omega(\Phi) + 3}\left(\kappa^{2}T - \frac{d\omega(\Phi)}{d\Phi}\left(\nabla\Phi\right)^{2} + \Phi\frac{dU}{d\Phi} - 2U\right),
\end{equation}
where $T$ is the trace of the matter energy-momentum tensor
\begin{equation}\label{}
T_{(m)\mu\nu} = - \frac{2}{\sqrt{-g}}\frac{\delta S_{m}}{\delta g^{\mu\nu}}.
\end{equation}
The field equation (\ref{eqJ}) can formally be written in the Einsteinian form,
\begin{equation}\label{gr}
R_{\mu\nu}- \frac{1}{2}g_{\mu\nu}R = \kappa^{2}T_{\mu\nu}
\end{equation}
with a total energy-momentum tensor
\begin{equation}\label{Ttot}
T_{\mu\nu} = T_{(m)\mu\nu} + T_{(x)\mu\nu},
\end{equation}
where $T_{(x)\mu\nu}$ is an effective energy-momentum tensor describing geometric ``matter" with
\begin{equation}\label{TxJ}
T_{(x)\mu\nu} \equiv \left(\frac{1}{\Phi}-1\right)T_{\mu\nu} + \frac{1}{\kappa^{2}\Phi}
\left[\frac{\omega(\Phi)}{\Phi}
\left(\partial_{\mu}\Phi\partial_{\nu}\Phi
- \frac{1}{2}g_{\mu\nu}\left(\nabla\Phi\right)^{2}\right)
+ \nabla_{\mu}\nabla_{\nu}\Phi- g_{\mu\nu}\Box\Phi
- \frac{1}{2}g_{\mu\nu}U\right].
\end{equation}
Because of the contracted Bianchi identities which imply $T^{\mu\nu}_{\ ;\nu} =0$ and because of the matter conservation law $T^{\mu\nu}_{(m) ;\nu} =0$ we have $T^{\mu\nu}_{(x);\nu} =0$ as well.
The scalar-tensor dynamics is thus mapped
onto an effective two-component model within GR, provided that $\Phi$ in (\ref{TxJ}) is a solution of (\ref{boxphi}).

\subsection{Effective fluid}
Assuming the existence of a timelike unit vector $u^{\mu}$, the timelike gradient of the scalar field, e.g.,  any tensor of the structure of (\ref{Ttot}) can be split according to
\begin{equation}\label{Tfgen}
T_{\mu\nu} = \rho u_{\mu}u_{\nu} + ph_{\mu\nu} + \Pi_{\mu\nu}
+ q_{\mu}u_{\nu} + q_{\nu}u_{\mu},
\end{equation}
where $u^{\mu}$ with $u^{\mu}u_{\mu}=-1$ characterizes a rest frame for the effective total cosmic substratum,
$h_{\mu\nu} = g_{\mu\nu} + u^{\mu}u^{\nu}$ is the spatial projection tensor and
\begin{equation}\label{projrho}
\rho = T_{\mu\nu}u_{}^{\mu}u_{}^{\nu}, \quad p  = \frac{1}{3}h^{\mu\nu}T_{\mu\nu},\quad
q_{\mu} = - h_{\mu}^{\nu}T_{\nu\sigma}u^{\sigma},\quad\Pi_{\mu\nu} = h_{(\mu}^{\sigma}h_{\nu)}^{\tau}T_{\sigma\tau} - \frac{1}{3}h_{\mu\nu} h^{\sigma\tau}
T_{\sigma\tau}
\end{equation}
with
$q_{\mu}u^{\mu} = \Pi_{\mu\nu}u^{\mu} = \Pi^{\mu}_{\mu} =0$.
Here, $\rho$ is the total energy density, $p$ is the total isotropic pressure,  $q_{\mu}$
is the total energy-flux vector and $\Pi_{\mu\nu}$  is the symmetric and trace-free anisotropic stress tensor.
The matter part $T_{(m)\mu\nu}$ is assumed to be
pressureless matter, i.e.,
\begin{equation}\label{Tm}
T_{(m)\mu\nu} = \rho_{m}u_{(m)\mu}u_{(m)\nu},
\end{equation}
where $u_{(m)}^{\mu}$ with $u_{(m)}^{\mu}u_{(m)\mu}=-1$ is the matter four-velocity which may be associated with an observer. In general, it will be different from the previously introduced timelike unit vector $u^{\mu}$.
The matter-energy density $\rho_{m}$ is
\begin{equation}\label{}
\rho_{m} = T_{(m)\mu\nu}u^{\mu}_{(m)}u^{\nu}_{(m)}.
\end{equation}
In the following we shall make use of the fluid structure (\ref{Tfgen}) to effectively describe the dynamics based on equations (\ref{gr}) and (\ref{Ttot}) together with (\ref{TxJ})  and (\ref{Tm}).

\section{The background model}
\label{background}
Omitting calculational details, we recall the basic results of the previously established scalar-tensor extension of the $\Lambda$CDM model \cite{WHW}.
Its main characteristic is an explicit analytical expression for the Hubble rate in which deviations from the standard model are described by a single constant parameter which also governs the effective scalar-field dynamics. The background relations are found through a specific solution of the effective fluid dynamics in the Einstein frame which subsequently is converted into the Jordan frame via a conformal transformation.
The result is the explicit expression
\begin{equation}\label{H2Phi}
\frac{H^{2}}{H_{0}^{2}} =
\frac{A\Omega_{m0}a^{-3}}{\Phi}
+ \left[1 - A\Omega_{m0}\right]\Phi,\qquad A\equiv \frac{\left(1 +3m \right)^{2}}{1 - m}
\end{equation}
for the Hubble rate \cite{WHW}.
Here, $\Phi$ is given by
\begin{equation}\label{Phivarphi+}
\Phi = a^{-\frac{6m}{1+3m}},
\end{equation}
where $a$ is the scale factor of the Robertson-Walker metric and $\Omega_{m0} \equiv \frac{\rho_{m0}}{\rho_{0}}$ denotes the present (subindex 0) matter fraction. The constant parameter $m$ describes deviations from the standard $\Lambda$CDM model which corresponds to the limit $m=0 \Rightarrow A=  \Phi =1$.
With (\ref{Phivarphi+}) we found an explicit expression for the scalar-field variable without having solved the basic scalar-field equation.
We have solved the system of energy-balance equations for a simple case \cite{WHW}.
This solution implies an explicit scale-factor dependence of $\Phi$ which not necessarily has to be a solution of the original scalar-field equation (\ref{boxphi}).
Instead, it obeys an alternative effective second-order equation with an alternative effective potential that does not necessarily coincide with $U$ \cite{WHW}.
The point here is that the dynamics on the level of the fluid energy densities does not require the exact solution of the scalar-field equation (\ref{boxphi}).
On the other hand, the specific features of our fluid dynamics imply the existence of the effective scalar field $\Phi$ of the form of  (\ref{Phivarphi+}).
Strictly speaking,  we are no longer in the framework of JBD theory, which, however, provided the motivation for our model. The background dynamics can be obtained without an exact solution of (\ref{boxphi}) since the geometrical field equations depend on the scalar field only through the energy-momentum tensor \cite{LAmend}. Not all details of the dynamics of the scalar field are required here.

Formula (\ref{H2Phi}) represents an explicit analytic solution for the Hubble rate of our $e_{\Phi}\Lambda$CDM model.
The scalar $\Phi$ modifies the cosmological dynamics compared
with the GR based $\Lambda$CDM model. The deviations  from the $\Lambda$CDM model are entirely encoded in the constant parameter $m$ which is supposed to be small. The appearance of the scalar $\Phi$ in the Hubble rate (\ref{H2Phi}) changes the relative contributions of matter and the DE equivalent compared with the $\Lambda$CDM model.
For $\Phi = 1$, equivalent to $m=0$, we recover the $\Lambda$CDM model.
For any $\Phi \neq 1$, equivalent to $m\neq 0$, the expression (\ref{H2Phi})  represents a testable, alternative model with presumably small deviations from the $\Lambda$CDM model.

The matter energy density obeys $\rho_{m} \propto a^{-3}$ and
the fractional matter contribution is
\begin{equation}\label{}
\Omega_{m} = \frac{\rho_{m}}{\rho} = \frac{\Omega_{m0}a^{-3}}{A\Phi^{-1}\Omega_{m0}a^{-3} + \left[1 - A\Omega_{m0}\right]\Phi}.
\end{equation}
The geometric ``matter" part contributes with $\Omega_{x} = 1 - \Omega_{m}$.
%For different values of $m$ the fractional abundances $\Omega_{m}$ and $\Omega_{x}$ are visualized in FIG.~1.
Postulating an effective conservation equation $\dot{\rho}_{x} + 3H\left(1+w_{x}\right)\rho_{x} =0$, where $\rho_{x} \equiv \rho - \rho_{m}$, this corresponds to an effective, time-varying EoS parameter $w_{x}$ of the geometric DE,
\begin{equation}\label{wtil-1}
w_{x}(a) = - 1 + \frac{\frac{2m}{1+3m}\left[1 - A\Omega_{m0}\right]\Phi + \Omega_{m0}a^{-3}\left[\frac{1+m}{1+3m}A \Phi^{-1} - 1\right]}{\left[1 - A\Omega_{m0}\right]\Phi + \Omega_{m0}a^{-3}
\left[A \Phi^{-1} - 1\right]}.
\end{equation}
For $m=0$ it reduces to the $\Lambda$CDM value $w_{x} = - 1$.
At high redshift one has
 \begin{equation}\label{}
w_{x} \approx - 1 + \frac{\left[\frac{1+m}{1+3m}A \Phi^{-1} - 1\right]}{\left[A \Phi^{-1} - 1\right]}
\qquad\qquad (a\ll 1).
\end{equation}
 This value may be close to zero, i.e., the geometric DE mimics dust in this limit,
but the effective energy density $\rho_{x}$ will be negative for $m>0$. It crosses $\rho_{x}=0$ in the redshift range $10\gtrsim z \gtrsim 4$ for  values of $m$  in the range $0.001 < m < 0.01$ \cite{WHW}.
This behavior reflects that fact that the x-component is very different from a conventional fluid.
The total EoS is well behaved throughout.
At the present time the effective EoS parameter is
\begin{equation}\label{}
w_{x} = - 1 + \frac{\frac{2m}{1+3m} +  3m\Omega_{m0}}{1 - \Omega_{m0}} \qquad\qquad (a=1).
\end{equation}
For small $|m|$ this remains in the vicinity of $w_{x} = - 1$.
In the far future $w_{x}$ approaches
\begin{equation}\label{}
w_{x} \approx - 1 + \frac{2m}{1+3m}  \qquad\qquad (a\gg 1).
\end{equation}

From a statistical analysis using Supernovae data,
differential age data of old galaxies that have evolved passively and baryon acoustic oscillations, we found a best-fit value \cite{WHW} of
$m= 0.004^{+0.011 (1\sigma) \,\, +0.017 (2\sigma)}_{-0.011 (1\sigma)\,\, -0.017 (2\sigma)}$.
This is compatible with the $\Lambda$CDM model but leaves also room for small deviations.
Even a very small non-vanishing value of $|m|$ is, however, expected to modify the standard scenario of structure formation.
In particular, it will affect the matter-growth function as will be demonstrated below.

\section{General perturbation dynamics}
\label{genpert}

We restrict ourselves to scalar metric perturbations, described by the line element
\begin{equation}
\mbox{d}s^{2} = - \left(1 + 2 \phi\right)\mbox{d}t^2 + 2 a^2
F_{,a }\mbox{d}t\mbox{d}x^{a} +
a^2\left[\left(1-2\psi\right)\delta _{ab} + 2E_{,ab} \right] \mbox{d}x^a\mbox{d}x^b \,.\label{ds}
\end{equation}
Denoting first-order variables by a hat symbol, the perturbed
time components of the four-velocities are
\begin{equation}
\hat{u}_{0} = \hat{u}^{0} = \hat{u}_{(m)}^{0}  = \frac{1}{2}\hat{g}_{00} = - \phi\ .
\label{u0}
\end{equation}
We define the (three-) scalar quantities $v$, and $v_{m}$  by
\begin{equation}
a^2\hat{u}^{m} + a^2F_{,m} = \hat{u}_{m} \equiv v_{,m},  \qquad \mathrm{and}\qquad
a^2\hat{u}^{a}_{(m)} + a^2F_{,a} = \hat{u}_{(m) a} \equiv v_{m,a},
\label{}
\end{equation}
respectively. The covariant divergence of the matter four-velocity at first order is
\begin{equation}
u^{\mu}_{(m);\mu} = \frac{1}{a^2}\left(\Delta v_{m} +\Delta \chi\right) -
3\dot{\psi} - 3 H\phi, \qquad \chi \equiv a^2\left(\dot{E} -F\right),
\label{Thetaexpl}
\end{equation}
where $\Delta$ is the three-dimensional Laplacian.
In this paper we focus on effects due to a non-vanishing anisotropic pressure and we ignore the heat flux from now on.
It is the anisotropic pressure which gives rise to the gravitational slip, i.e., to a dynamics with $\phi\neq\psi$.
The perturbed components of the total energy-momentum tensor then are
\begin{equation}\label{}
\hat{T}_{0}^{0} = \hat{\rho} = \hat{\rho}_{x} + \hat{\rho}_{m},\quad \hat{T}_{a}^{0} = \left(\rho + p\right)\hat{u}_{a}, \quad \hat{T}_{a}^{b}  = \hat{p}\delta_{a}^{b} + \Pi_{a}^{b}.
\end{equation}
The scalar part $\Pi$ of  $\Pi_{ab}$ is defined by
\begin{equation}\label{}
\Pi_{ab} = \left(\partial_{a}\partial_{b} - \frac{1}{3}\delta_{ab}\Delta\right)\Pi.
\end{equation}
Non-vanishing anisotropic pressure is a typical feature in extended theories of gravity \cite{bertschinger,bean,hojjati,Silvestri,LAmend}.
With $\delta = \frac{\hat{\rho}}{\rho}$ and $\psi^{\chi} = \psi + H\chi$
the $00$ and $0a$ field equations provide us with the Poisson equation (transformed into the $\mathbf{k}$ space via
$\Delta \rightarrow -k^{2}$)
\begin{equation}\label{Newtonsim}
\frac{k^{2}}{a^{2}}\psi^{\chi} = -4 \pi G\rho\delta^{c},
\end{equation}
where
$\delta^{c} = \delta -3H\left(1+w\right)v$ is the total comoving density perturbation and
$w=\frac{p}{\rho}$ is the total EoS parameter.

\section{Modeling the anisotropic pressure}
\label{aipressure}

The total energy-density perturbation that determines the potential $\psi$ can, in principle, be obtained from the exact scalar-field  dynamics (\ref{eqJ}) and (\ref{boxphi}).
At first order one finds (see, e.g., \cite{HwangNoh2005,chibayam})
\begin{eqnarray}
\label{hrhoexact}
    \hat{\rho} &=&  \frac{\hat{\rho}_{m}}{\Phi}+ \frac{1}{\kappa^{2}\Phi}\left\{\frac{1}{2}\frac{\omega}{\Phi}
  \left[2\frac{\partial\Phi}{\partial t}\hat{\Phi}_{,0}+ \left(\frac{d\omega}{d\Phi}\frac{\hat{\Phi}}{\omega} - \frac{\hat{\Phi}}{\Phi}-2\phi\right)\left(\frac{\partial\Phi}{\partial t}\right)^{2}\right]\right.\nonumber  \\
  && \left.  - 3H\left(\frac{\partial\Phi}{\partial t} + H\hat{\Phi}\right)
  + 3 \frac{\partial\Phi}{\partial t}\left(2H\phi + \dot{\psi}\right) + \frac{1}{2}\frac{dU}{d\Phi}\hat{\Phi}
  + \frac{1}{a^{2}}\delta_{ab}\hat{\Phi}_{,ab}
 \right\}.
\end{eqnarray}
Analogous relations can be obtained for the perturbations of the other fluid quantities in (\ref{projrho}) as well as for their counterparts of the individual components, {in particular also for the anisotropic pressure (\cite{HwangNoh2005,chibayam}).

On the other hand, the comoving energy-density perturbation $\delta^{c}$ obeys the fluid-dynamical conservation equations following from  $T^{\mu\nu}_{\ ;\nu} =0$ at first order. In general, in the presence of pressure perturbations, there will be no closed equation for $\delta^{c}$.

From the total fluid conservation dynamics together with the Raychaudhuri equation for the expansion scalar one obtains (cf. \cite{baVDF,ricci,alonso})
\begin{eqnarray}\label{epsprpr}
&&\delta^{c\prime\prime} + \left(\frac{3}{2} - \frac{15}{2}\frac{p}{\rho} + 3\frac{p^{\prime}}{\rho^{\prime}}\right)\frac{\delta^{c\prime}}{a}
 - \left[\frac{3}{2} + 12\frac{p}{\rho} - \frac{9}{2}\frac{p^{2}}{\rho^{2}} - 9\frac{p^{\prime}}{\rho^{\prime}}\right]\frac{\delta^{c}}{a^{2}}
 + \frac{1}{a^2 H^{2}}\frac{k^{2}}{a^{2}}\frac{\hat{p}^{c}}{\rho}
 \nonumber\\
&&\qquad\qquad\qquad + \frac{2}{a}\frac{k^{2}\Pi^{\prime}}{a^{2}\rho}
- \left[-3\left(1-\frac{p}{\rho}+2\frac{p^{\prime}}{\rho^{\prime}}\right)\frac{k^{2}}{a^{2}H^{2}}
    +\frac{2}{3} \frac{k^{4}}{a^{4}H^{4}}
    \right]\frac{H^{2}}{a^{2}}\frac{\Pi}{\rho}
=0,
\end{eqnarray}
where the prime denotes a derivative with respect to the scale factor and $\hat{p}^{c} =\hat{p} + \dot{p}v $. Equation (\ref{epsprpr}) generalizes the corresponding perfect-fluid equation which is recovered for $\Pi =0$. Even for the perfect-fluid case it is generally not a closed equation for $\delta^{c}$ unless there exists an EoS $p=p(\rho)$. Via the isotropic pressure perturbations $\hat{p}^{c}$ the dynamics of  the density perturbation $\delta^{c}$ in a multi-component system is coupled to entropy-type relative perturbations for the components. An additional equation for these relative perturbations has to be established to obtain a closed system of equations for the two-component system. The presence of anisotropic pressures adds to the complexity of the coupling.

With the help of the (gauge-invariant) quantities
\begin{equation}\label{}
 \hat{\rho}_{m}^{c} = \hat{\rho}_{m} + \dot{\rho}_{m}v,\quad
\delta^{c}_{m} = \frac{\hat{\rho}_{m}^{c}}{\rho_{m}},
\end{equation}
we define the relative density perturbations
\begin{equation}\label{defSm}
S_{m} \equiv \frac{\delta^{c}}{1+w} - \delta^{c}_{m},\qquad w= \frac{p}{\rho},
\end{equation}
as the difference between total and pure matter perturbations.
The perturbations of the pressure are the sum of an adiabatic part $\frac{\dot{p}}{\dot{\rho}}\hat{\rho}^{c}$ and a non-adiabatic contribution $\hat{p}_{nad}$,
\begin{equation}
\hat{p}^{c} = \frac{\dot{p}}{\dot{\rho}}\hat{\rho}^{c} + \hat{p}_{nad}.
\label{pc}
\end{equation}
For our configuration the non-adiabatic part explicitly becomes
\begin{equation}
\hat{p}_{nad}
 =  \hat{p}_{x}^{c}-
\frac{\dot{p}_x}{\dot{\rho}_x}\hat{\rho}_{x}^{c}
+ \frac{\rho_{m} \left(\rho_{x} + p_{x}\right)}{\left(\rho +
p\right)} \frac{\dot{p}_x}{\dot{\rho}_x}
\left[\frac{\hat{\rho}_{x}^{c}}{\rho_{x} + p_{x}} -
\frac{\hat{\rho}_{m}^{c}}{\rho_{m}} \right].
\label{pna}
\end{equation}
Here, the combination $\hat{p}_{x}^{c}-
\frac{\dot{p}_x}{\dot{\rho}_x}\hat{\rho}_{x}^{c}$ accounts for the intrinsic non-adiabatic perturbations of the x component while the last term appears due to the two-component nature of the cosmic medium. As a consequence, the fluid as a whole is non-adiabatic even if each of its components is adiabatic on its own.
With $\hat{\rho}_{x}^{c} = \hat{\rho}^{c} - \hat{\rho}_{m}^{c}$ in the last term on the right-hand side of
(\ref{pna}) we may write
\begin{equation}\label{}
\frac{\hat{\rho}_{x}^{c}}{\rho_{x} + p_{x}} - \frac{\hat{\rho}_{m}^{c}}{\rho_{m}}
= \frac{\rho + p}{\rho_{x} + p_{x}} \left(\frac{\delta^{c}}{1+w} - \delta^{c}_{m} \right) = \frac{\rho + p}{\rho_{x} + p_{x}}S_{m}.
\end{equation}
Then the non-adiabatic pressure perturbations are
\begin{equation}
\hat{p}_{nad}
 =  \hat{p}_{x}^{c}-
\frac{\dot{p}_x}{\dot{\rho}_x}\hat{\rho}_{x}^{c}
+ \rho_{m}\frac{\dot{p}_x}{\dot{\rho}_x}
S_{m}.
\label{pna2}
\end{equation}
This implies that through the  pressure perturbations  the dynamics of the  total energy-density perturbation $\delta^{c}$ is coupled to the dynamics of $S_{m}$
(cf. (\ref{epsprpr})).
To obtain the dynamics of $S_{m}$ we have to combine the conservation equations for the total medium with those for the matter component. The result is
\begin{eqnarray}
S_{m}^{\prime\prime} +  \frac{3}{2}\left(1-\frac{p}{\rho}\right)\frac{S_{m}^{\prime}}{a} + \frac{k^{2}}{a^{2}H^{2}}\frac{\hat{p}^{c}}{a^{2}\left(\rho +p\right)}
+ \frac{3}{a}\frac{\hat{p}^{\prime}_{nad}}{\rho +p} + 9\left(\frac{7}{6}+\frac{p^{\prime}}{\rho^{\prime}}  - \frac{1}{2}\frac{p}{\rho}\right)
\frac{\hat{p}_{nad}}{a^{2}\left(\rho +p\right)}&&\nonumber  \\
-\frac{2}{3}\frac{1}{a^{2}H^{2}}\frac{k^{4}\Pi}{a^{4}\left(\rho +p\right)}
&=& 0.
\label{sprpr}
\end{eqnarray}
As in Eq.~(\ref{epsprpr}), neither the isotropic pressure perturbations $\hat{p}^{c}$ nor the anisotropic pressure $\Pi$ are specified.
In the perfect-fluid case, i.e., in the absence of anisotropic pressures, the pressure perturbations are known to induce a coupling between equations (\ref{epsprpr}) and (\ref{sprpr}) (cf. \cite{baVDF,ricci,alonso}).
Since the matter is pressureless, the total
(generally non-adiabatic) isotropic pressure perturbation coincides with the perturbation of the effective pressure of component $x$ which in the rest frame (we ignore here the difference between the total cosmic rest frame and the rest frame of the x component)
is related to the rest-frame energy-density perturbation $\hat{\rho}_{x}^{c}$
via the square of the sound speed $c_{x}^{2}$ according to
\begin{equation}\label{}
  \hat{p}_{x}^{c} = \hat{p}^{c}= c_{x}^{2}\hat{\rho}_{x}^{c}.
\end{equation}
The sound speed square is a free parameter which is expected to assume a value between zero (non-relativistic matter) and one (scalar field).
With
\begin{equation}
\hat{\rho}_{x} = \hat{\rho} - \hat{\rho}_{m} = \rho\delta - \rho_{m}\delta_{m}\quad \mathrm{and} \quad \delta_{m} = \frac{\delta}{1+w} - S_{m}
\nonumber
\end{equation}
the pressure perturbation is equivalent to
\begin{equation}\label{pcfin}
\hat{p}^{c} = c_{x}^{2}\frac{\rho_{x} + p_{x}}{\rho + p}\rho\delta^{c} + c_{x}^{2}\rho_{m}S_{m},
\end{equation}
i.e., it is determined both by the total energy-density perturbations $\delta^{c}$ and by the relative perturbations $S_{m}$.
The first term in (\ref{pcfin}) represents the adiabatic part of the (isotropic) pressure perturbation, the second term accounts for the non-adiabaticity.
Via the pressure perturbations $\hat{p}^{c}$ the equations (\ref{epsprpr}) and (\ref{sprpr}) for $\delta^{c}$ and $S_{m}$, respectively, are coupled to each other.
The new feature here is the appearance of
a so far unspecified anisotropic pressure $\Pi$.
Guided by the common practice  to describe  isotropic pressure perturbations through a phenomenological sound speed parameter which leads to (\ref{pcfin}), the idea is to determine the anisotropic pressure $\Pi$ in a similar way in terms of our basic variables $\delta^{c}$ and $S_{m}$ and thus to close the system (\ref{epsprpr}) and (\ref{sprpr}).
The most natural way to do this is to assume the so far unspecified anisotropic pressure $\Pi$ to be given in terms of a combination of the two independent perturbation variables $\delta^{c}$  and  $S_{m}$
 as well. This  parallels relation (\ref{pcfin}) for the isotropic pressure.
We write for the anisotropic pressure
\begin{equation}\label{ansPi}
\frac{8\pi G}{3}\Pi= \mu \delta^{c} + \nu \frac{\rho_{m}}{\rho}S_{m}
\end{equation}
with phenomenological coefficients $\mu$ and $\nu$.
This ansatz is similar to relation (\ref{pcfin}) by which the effective sound speed of the DE component is introduced. The coefficients $\mu$ and $\nu$ play a similar r\^{o}le in (\ref{ansPi}) as the sound speed square does in (\ref{pcfin}).
Moreover, they quantify the difference between the gravitational potentials $\psi$ and $\phi$ and thus may provide a measure for deviations from the GR based standard model.
With the ansatz (\ref{ansPi}) the system for $\delta^{c}$  and  $S_{m}$ is closed.
%Relation (\ref{ansPi}) is entirely phenomenological but it will allow us to put limits on the magnitude of the %anisotropic pressure.
Like the sound speed parameter $c_{x}$ the coefficients $\mu$ and $\nu$ are expected to be calculable from  an underlying microscopic theory. Here they are treated as phenomenological fluid quantities.
We recall that within JBD theory the anisotropic pressure $\Pi$ is given in terms of the perturbed scalar field (cf. \cite{HwangNoh2005}). Our approach, although inspired by JBD theory, in particular with respect to the background dynamics, does not rely on JBD perturbation theory. It uses the fluid dynamical conservation equations which do not require the explicit
scalar-field dynamics. Our basic set of perturbation equations is valid for any homogeneous and isotropic, spatially flat background, not only for the JBD-inspired background solution (\ref{H2Phi}) which will be used for our data analysis in Sec. \ref{data}. 
Our scheme should apply to any theory which can formally be transformed to an Einstein-type theory with an effective energy-momentum tensor. Different theories should  lead to different values of the parameters $c_{x}$, $\mu$ and $\nu$. Here we restrict ourselves to the impact that (an-)isotropic pressure perturbations may have on the matter growth rate on a purely phenomenological basis.
For phenomenological relations between anisotropic stresses and energy-density perturbations in different contexts see,  e.g.,  \cite{kunzsapone07,cardona,blas}.
Our setup has also similarities with the parametrization of perturbations via equations of state,  put forward in  \cite{battye,battye13,battye16} on the basis of a rather general scalar-field Lagrangian.
In this approach, after elimination
of internal degrees of freedom, equations of state for the
entropy perturbation and the anisotropic stress in terms of perturbations of the density, the velocity and
the metric perturbations are introduced in order to obtain closed perturbation equations.
What is different among others is the choice of basic variables. In particular, the entropy perturbation is one of the two basic dynamical quantities in our context and neither velocity components nor metric functions do appear explicitly
in the system of equations.
An imperfect fluid description of scalar-tensor theories has been recently performed also in \cite{faraoni}.

The anisotropic pressure  $\Pi$ is known to give rise to a difference in the potentials $\phi$ and $\psi$.
From the space-space field equation it follows that (in the longitudinal gauge $\chi =0$)
\begin{equation}\label{psi-phi=}
\psi - \phi = 8\pi G\Pi.
\end{equation}
Combination with (\ref{ansPi}) yields
\begin{equation}\label{phipsi}
\phi = \left[1+2\frac{k^{2}}{a^{2} H^{2}}\left(\mu + \nu \frac{\rho_{m}}{\rho}\frac{S_{m}}{\delta^{c}}\right)\right]\psi.
\end{equation}
The difference in the gravitational potentials $\phi$ and $\psi$ and
the consequences for the matter growth (see below) have been widely used to discriminate modified gravity from GR with the help of different parametrizations for
the gravitational slip $\frac{\psi}{\phi}$ \cite{boisseau,tsujikawa07,bertschinger,gannouji,song09,bean,hojjati,Silvestri,steigerwald14,piazza14}.
We emphasize, however, that (\ref{phipsi}) is not just a parametrization of the gravitational slip $\frac{\psi}{\phi}$. Knowledge of this ratio of the potentials requires the solution of the entire perturbation dynamics of the coupled system for $\delta^{c}$ and $S_{m}$.

\section{Matter perturbations}
\label{matter}

The matter perturbations $\delta_{m}^{c}$ are given in terms of $\delta^{c}$ and $S_{m}$ by
\begin{equation}\label{deltamDS}
\delta_{m}^{c} = \frac{\delta^{c}}{1+w} -S_{m}.
\end{equation}
Formally, this is an identity following from (\ref{defSm}).
Since the entire dynamics is described through $\delta^{c}$  and  $S_{m}$, relation (\ref{deltamDS}) can be used to find the matter perturbations from the solutions of the system (\ref{epsprpr}) and (\ref{sprpr}).
The aim of this paper is, however, to present a substantially simplified manner to determine the matter perturbations which bypasses the explicit use of the system (\ref{epsprpr}) and (\ref{sprpr}).
To this purpose we realize that in terms of the energy densities $\delta^{c}_{m}$ and $\delta^{c}_{x}$ of the components the total density perturbation $\delta^{c}$ is given by
\begin{equation}\label{}
\delta^{c} = \frac{\rho_{m}}{\rho}\delta^{c}_{m} + \frac{\rho_{x}}{\rho}\delta^{c}_{x}.
\end{equation}
A drastic simplification is achieved if we assume a proportionality with a (generally scale dependent) factor $y$ between $\delta^{c}_{x}$ and $\delta^{c}_{m}$,
\begin{equation}\label{y}
\delta^{c}_{x} = y \delta^{c}_{m}.
\end{equation}
Then
\begin{equation}\label{deltay}
\delta^{c} = \left(\frac{\rho_{m}}{\rho} +y \frac{\rho_{x}}{\rho}\right)\delta^{c}_{m} =
 \left(\Omega_{m} +y \left(1 - \Omega_{m}\right)\right)\delta^{c}_{m}, \qquad \Omega_{m} = \frac{\rho_{m}}{\rho}
\end{equation}
and
\begin{equation}\label{}
S_{m} = -  \left(\frac{\rho_{x}+p_{x}}{\rho +p} - y\frac{\rho_{x}}{\rho +p}\right)\delta^{c}_{m}.
\end{equation}
Now the relevant ratio $\frac{\rho_{m}}{\rho}\frac{S_{m}}{\delta^{c}}$ in (\ref{phipsi}) reduces to
\begin{equation}\label{}
\frac{\rho_{m}}{\rho}\frac{S_{m}}{\delta^{c}}=  -  \frac{\rho_{x}+p_{x}}{\rho +p}
+ y\frac{1 - \Omega_{m}}{\Omega_{m} + y\left(1 - \Omega_{m}\right)}
\end{equation}
with the help of which the ratio $\frac{\phi}{\psi}$ becomes
\begin{equation}\label{slip}
\frac{\phi}{\psi} = 1+2\frac{k^{2}}{a^{2} H^{2}}\left[\mu - \nu\left( \frac{\rho_{x}+p_{x}}{\rho +p} - y \frac{1 - \Omega_{m}}{\Omega_{m} + y\left(1 - \Omega_{m}\right)}\right)\right].
\end{equation}
The parameter $y$ quantifies the r\^{o}le of perturbations of the geometric DE component. For $y=0$ there are no DE perturbations.  If, moreover, $p_{x} =- \rho_{x}$, which corresponds to $m=0$, the $\Lambda$CDM case with $\phi = \psi$  is recovered.
In many studies DE perturbations were assumed to be negligible, at least on sub-horizon scales. Explicit calculation confirmed this for special cases \cite{pertsaulo}.
But in any consistent theory of dynamical DE these perturbations naturally appear and have to be taken into account
\cite{BuenoS,SaltasKunz,DossettIshak,amendola13,silveirawaga,kunzsapone07,abramo07,saponekunz09,Park-Hwang,song,nesseris15,basilakos15,nesseris17}.

With (\ref{slip}), based on the assumption (\ref{y}), the matter perturbations can be calculated without explicitly solving the coupled general system of perturbation equations. Starting point is the
first-order energy conservation equation for matter,
\begin{equation}\label{ebB1}
\dot{\hat{\rho}}_{m} + \dot{\rho}_{m}\hat{u}^{0} + \hat{\Theta}_{m}\rho_{m} + \Theta\hat{\rho}_{m} = 0,
\end{equation}
or
\begin{equation}\label{ebB2}
\dot{\delta}_{m} + 3H\phi + \hat{\Theta}_{m} = 0,
\end{equation}
where the perturbation of $\Theta_{m} \equiv u^{\alpha}_{m;\alpha}$ is explicitly given by
(\ref{Thetaexpl}).
Then the matter energy conservation becomes
\begin{equation}\label{dmconserv}
\dot{\delta}_{m}  - 3\dot{\psi} -\frac{k^{2}}{a^2}\left(v_{m} +\chi\right) =0.
\end{equation}
The first-order momentum conservation of the matter component yields
\begin{equation}\label{mmconserv}
\dot{v}_{m} = - \phi .
\end{equation}
Differentiating the energy conservation equation (\ref{dmconserv}), using the momentum conservation (\ref{mmconserv}) and then the energy conservation again,  results in
\begin{equation}\label{}
\ddot{\delta}_{m} + 2H\dot{\delta}_{m} + \frac{k^{2}}{a^2}\left(\phi +\dot{\chi}\right) = 3\left(\ddot{\psi} + 2H\dot{\psi}\right).
\end{equation}
This is still exact at first order.
With the gauge $\chi =0$ and in the  quasi-static
sub-horizon approximation one has
\begin{equation}\label{}
\ddot{\delta}_{m} + 2H\dot{\delta}_{m} + \frac{k^{2}}{a^2}\phi= 0.
\end{equation}
On sub-horizon scales we may drop the superscript c since gauge issues are irrelevant here.
With the assumption
\begin{equation}\label{}
\psi = \gamma\phi
\end{equation}
and  with (\ref{Newtonsim})
we have
\begin{equation}\label{}
\ddot{\delta}_{m} + 2H\dot{\delta}_{m}  -4\pi \frac{G}{\gamma} \rho \delta= 0.
\end{equation}
With our relation (\ref{deltay}) this results in a closed equation for $\delta_{m}$,
\begin{equation}\label{dGeff}
\ddot{\delta}_{m} + 2H\dot{\delta}_{m}  -4\pi G_{eff} \rho \Omega_{m}\delta_{m}= 0,
\end{equation}
where, according to  (\ref{deltay}) and (\ref{slip}) the effective gravitational ``constant" $G_{eff}$ is
\begin{equation}\label{Ge}
G_{eff} = \left[1 +y \frac{\left(1 - \Omega_{m}\right)}{\Omega_{m}}\right]
\left[1+2\frac{k^{2}}{a^{2} H^{2}}\left(\mu - \nu\left(\frac{\rho_{x}+p_{x}}{\rho +p} - y \frac{1 - \Omega_{m}}{\Omega_{m} + y\left(1 - \Omega_{m}\right)}\right)\right)\right]G.
\end{equation}
The effective gravitational coupling is explicitly scale dependent. The factor $y$ may introduce a further scale dependence.  Changing to the variable $a$, we have
\begin{equation}\label{dmprpr}
\delta_{m}^{\prime\prime} + \frac{3}{2}\left(1-w\right)\frac{\delta_{m}^{\prime}}{a}
-\frac{3}{2}\frac{G_{eff}}{G}\Omega_{m}\frac{\delta_{m}}{a^{2}} =0.
\end{equation}
In the absence of anisotropic stresses ($\mu=\nu =0$) a positive $y$ enhances $G_{eff}$, a negative $y$ diminishes it. A negative $\mu$ tends to reduce $G_{eff}$. For $\frac{k^{2}}{a^{2} H^{2}}|\mu| > \frac{1}{2}$ it may even become negative. The impact of the $\nu$ term is expected to be small since $\rho_{x}+p_{x}$ is close to zero.
For scales of the order of $k \approx 0.1 h {\mathrm{Mpc}^{-1}}$ the present value of the factor $\frac{k^{2}}{a^{2} H^{2}}$ is (restoring the units appropriately) $\frac{k^{2}c^{2}}{H_{0}^{2}} \approx 1.8\cdot 10^{5}$.
Consequently, an anisotropic pressure with $|\mu|$ of the order of $|\mu| \approx 10^{-6}$ should have a noticeable impact on the effective gravitational coupling.
We mention that our simplified treatment leading to equation (\ref{dGeff}) implies that the anisotropic pressure enters only via the $G_{eff}$ factor while the damping term remains unaffected. The structure of (\ref{epsprpr}) suggests that in the full theory the coefficient multiplying $\dot{\delta}_{m}$ will have contributions proportional to $\mu$ and $\nu$ as well.

\section{Data analysis}
\label{data}

Equation (\ref{dmprpr}) describes how the late-time accelerated expansion tends to smooth the matter perturbations. Current galaxy surveys provide observational data for the combination $f \sigma_8$ where the linear growth rate $f$ is defined as $f= {\rm d\, ln}\, \delta_m / {\rm d\, ln}\, a$ and $\sigma_8$ is the root-mean-square mass fluctuation in spheres with radius $8h^{-1}\mathrm{Mpc}$  \cite{song09}.
In terms of the quantity $f$, equation (\ref{dmprpr}) can be recast in the form
\begin{equation}\label{dfdlna}
\frac{d f}{d\ln a} + f^{2} + \frac{1}{2}\left(1-3w\right) = \frac{3}{2}\frac{G_{eff}}{G}\Omega_{m}.
\end{equation}
This scale-dependent equation is valid in the linear regime, it is expected to  break down for non-linear modes. We shall trace the evolution of matter overdensities from a moment deep in the matter dominated epoch until today. To stay in the linear regime  we fix the value $k = 0.1 h {\mathrm{Mpc}^{-1}}$ in our analysis. With this assumption the temporal evolution of the normalization $\sigma_8$ follows the $\delta_m$ amplitude such that \cite{nesseris08}
\begin{eqnarray}
\sigma _8(z)=\frac{\delta_{m} (z)}{\delta (z=0)}\sigma _8(z=0)
\end{eqnarray}
and
\begin{eqnarray}
f\sigma_8 (z)= -(1+z)\frac{\sigma_8(z=0)}{\delta_{m}(z=0)}\frac{d}{dz}\delta_{m}(z).
\end{eqnarray}
Unfortunately, the amount of currently available $f(z) \sigma_8(z)$ data is not representative (a couple of dozens) for a statistical analysis. Also, the variance in the data is still high, lowering the confidence in any statistical {\it a posteriori} result. What can be done, however, is to assess the impact of changing the model parameter values in comparison to the $\Lambda$CDM reference curve.
For the matter distribution variance today we assume $\sigma_8(z=0)=0.8$ which is compatible with standard fiducial cosmology as determined by the Planck satellite. This value is biased with respect to the variance in the distribution of galaxies but the combination $f \sigma_8$ is independent of the bias factor \cite{song09}.
In our analysis we use the 18 `Gold' growth set data points of RSD measurements of $f \sigma_8$
 listed in \cite{nesseris17}.
 In the following we study the influence of different parameter combinations on the growth function and on the effective gravitational ``constant". We start by considering the impact of each of the parameters separately.
It is from this point on that we make explicit use of the background solution (\ref{H2Phi}).

The left panel of FIG.~1 shows the redshift dependence of the growth function $f\sigma_8 (z)$ if only the background dynamics is changed compared with that of the standard model, i.e., anisotropic pressures and DE perturbations are absent ($\mu=\nu=y =0$). Positive values of $m$ lower the model predictions with respect to $\Lambda$CDM. Negative values result in a shift in the opposite direction for $z\lesssim 1$. For all $m$ values the largest changes appear at $z\approx 0.5$ which seems to be the optimal redshift range for seeking background effects in the $f \sigma_8$ observable.
It is worth noting that if only $m\neq 0$ but $\mu=\nu=y =0$ the effective gravitational constant reduces to its Newtonian value, i.e., $G_{eff}=G$.
The right panel of  FIG.~1 conjoins the evolutionary tracks of $H(z)/H_{0}$ and $f\sigma_8 (z)$ as suggested in \cite{Linder16}. These conjoined tracks are particularly useful for the varying $m$
case since both background and perturbations are affected simultaneously.
In FIG.~2 we depict the impact of DE perturbation on $f\sigma_8 (z)$ for the $\Lambda$CDM background value $m=0$
in the absence of anisotropic pressures ($\mu=\nu= 0$). The different curves correspond to different fractional contributions $y$ from DE perturbations.
There is a tendency to heighten  the pure $\Lambda$CDM curve at small redshift for positive values of $y$.
For $y<0$ the curve is lowered.
The corresponding effective gravitational ``constant" is shown in the right panel. If the DE perturbation parameter $y$ becomes of the order of one, the deviations from the standard model become unacceptably large. This implies that on the scale in question the DE perturbations are at least one order of magnitude smaller than the matter perturbations.
FIG.~3 demonstrates the influence of the anisotropy parameter $\mu$ on $f\sigma_8 (z)$ for the fixed background value $m=0$ in case there is no coupling of $\Pi$ to the relative perturbations ($\nu = 0$) and DE perturbations are absent ($y=0$). The right panel shows the corresponding influence of $\mu$ on the effective gravitational ``constant".
FIG.~4 visualizes the influence of the anisotropy parameter $\nu$ on $f\sigma_8 (z)$
and on the effective gravitational ``constant" for the fixed background value $m=0$ and
for $\mu = y=0$. The influence of $\nu$ separately on $f\sigma_8 (z)$ is weaker than the influence of a separate $\mu$ of the same order (cf. FIG.~3). Moreover, it acts in the opposite direction.
FIG.~5 depicts the relevance of the cross term that involves both $\nu$ and $y$ in the expression (\ref{Ge}) for $G_{eff}$.
Different from the situation of FIG.~2, in this term positive values of $y$ lower the curve for $f\sigma_8$ (for $z\lesssim 1$), corresponding to a reduced effective gravitational ``constant".  \\

%%%%%%%%%%%%%%%%%%%%%%%%%%%%%%%%%%%%%%%%%%%%%%%%%%%%%%%%%%%

\begin{figure}[h!]
\begin{center}
\includegraphics[width=0.45\textwidth]{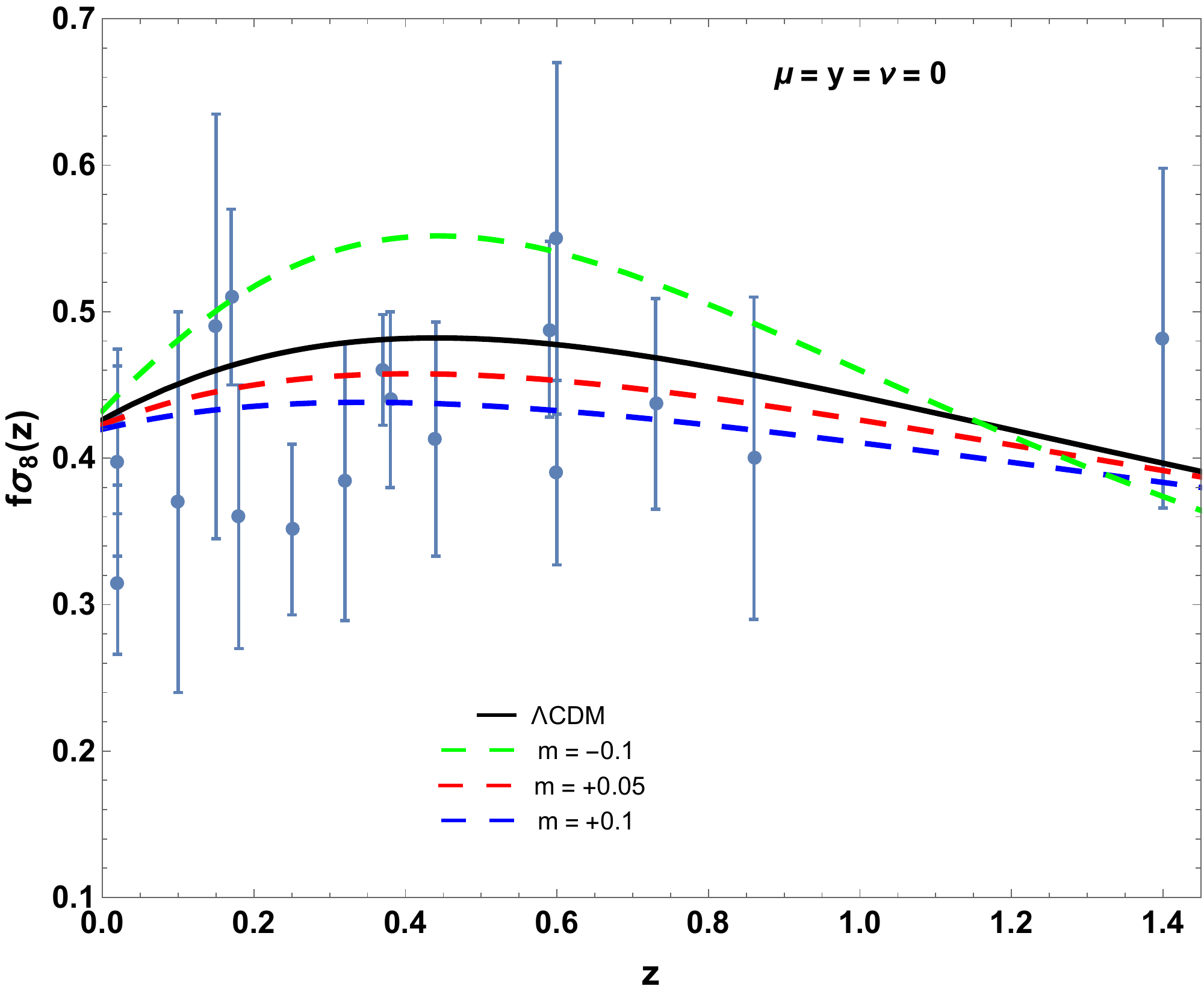}
\hspace{1cm}
\includegraphics[width=0.45\textwidth]{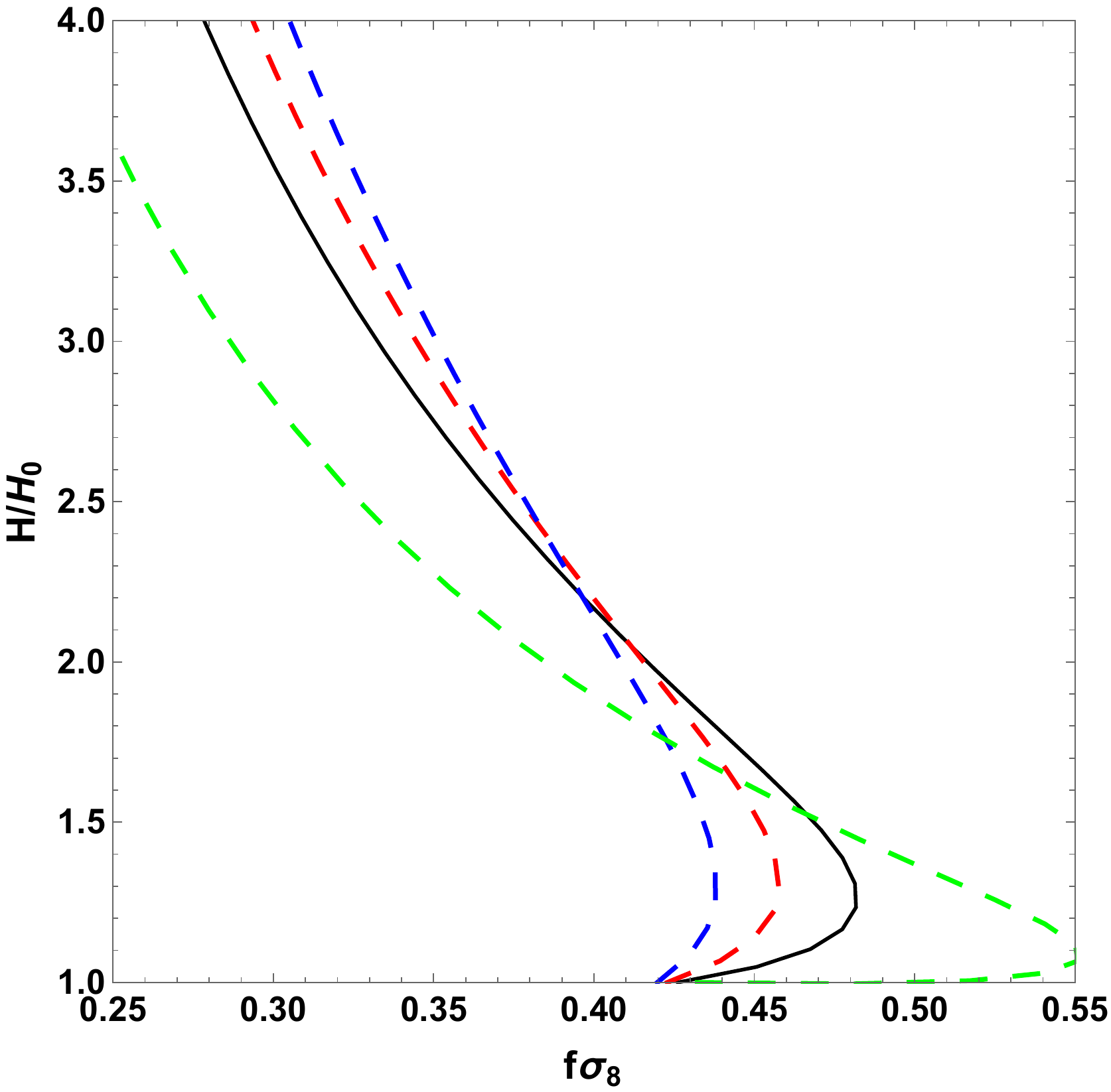}
\label{}	
\caption{Left panel: dependence of $f\sigma_8 (z)$ on $z$ if only the background dynamics is changed ($\mu=\nu=y =0$). Positive values of $m$ lower the $\Lambda$CDM prediction for $z\lesssim 1$, negative values result in a shift in the opposite direction. The largest changes appear at $z\approx 0.5$.
The right panel shows the corresponding conjoined evolutionary tracks of $H(z)/H_{0}$ and $f\sigma_8 (z)$. }
\end{center}
\end{figure}

\begin{figure}[h!]
\begin{center}
\includegraphics[width=0.45\textwidth]{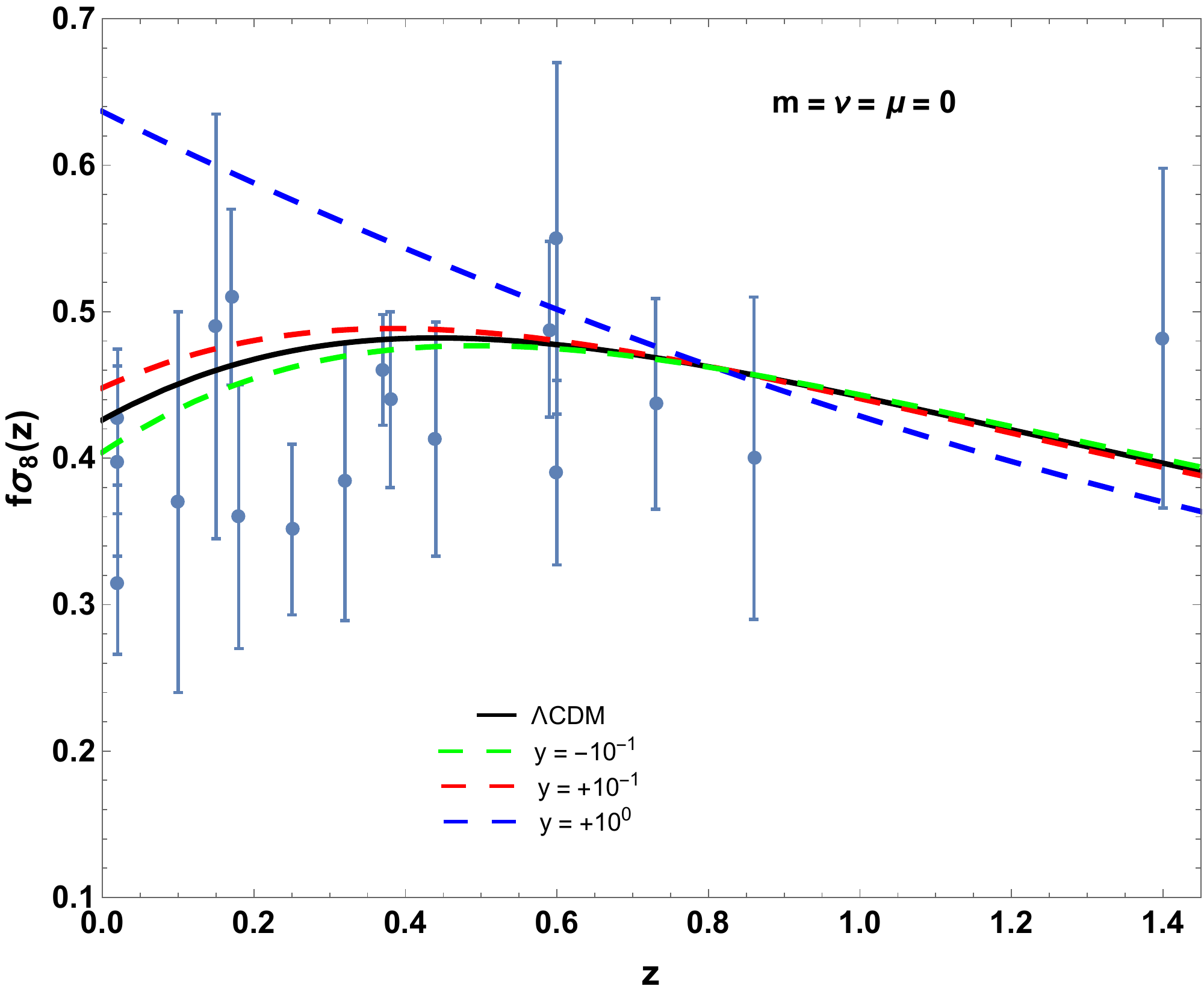}
\hspace{1cm}
\includegraphics[width=0.45\textwidth]{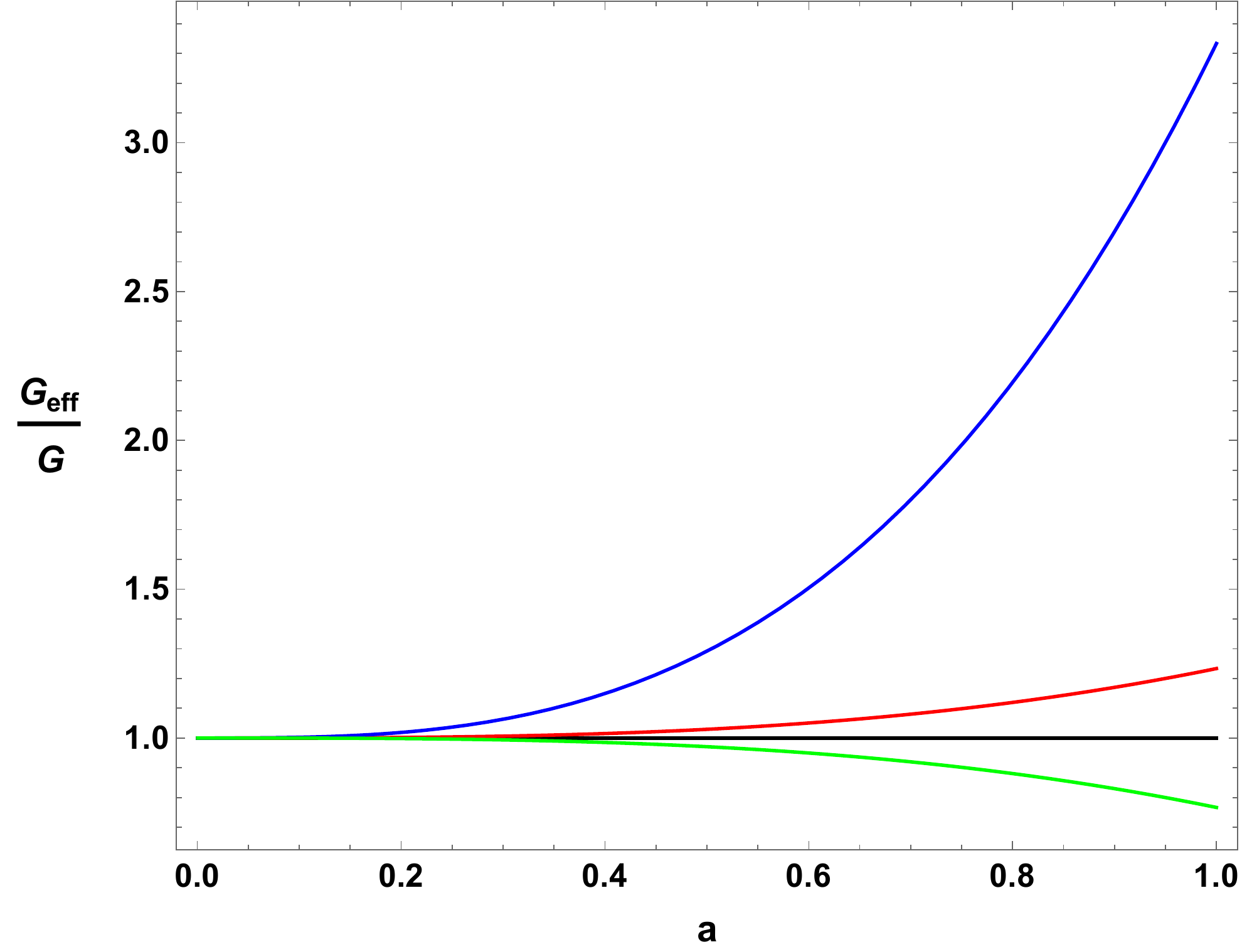}
\label{}	
\caption{Left panel: dependence of $f\sigma_8 (z)$ on $z$ for the $\Lambda$CDM background value $m=0$ with $\mu=\nu= 0$. The different curves correspond to different contributions from DE perturbations.
For $y>0$ there is a tendency to heighten the pure $\Lambda$CDM curve at small redshift. For $y<0$ the shift is to lower values for $z\lesssim 1$. Right panel:
effective gravitational ``constant". If $y$ is of the order of one, i.e., if DE perturbations are of the same order as the matter perturbations, the deviations become unacceptably large.}
\end{center}
\end{figure}

\begin{figure}[h!]
\begin{center}
\includegraphics[width=0.45\textwidth]{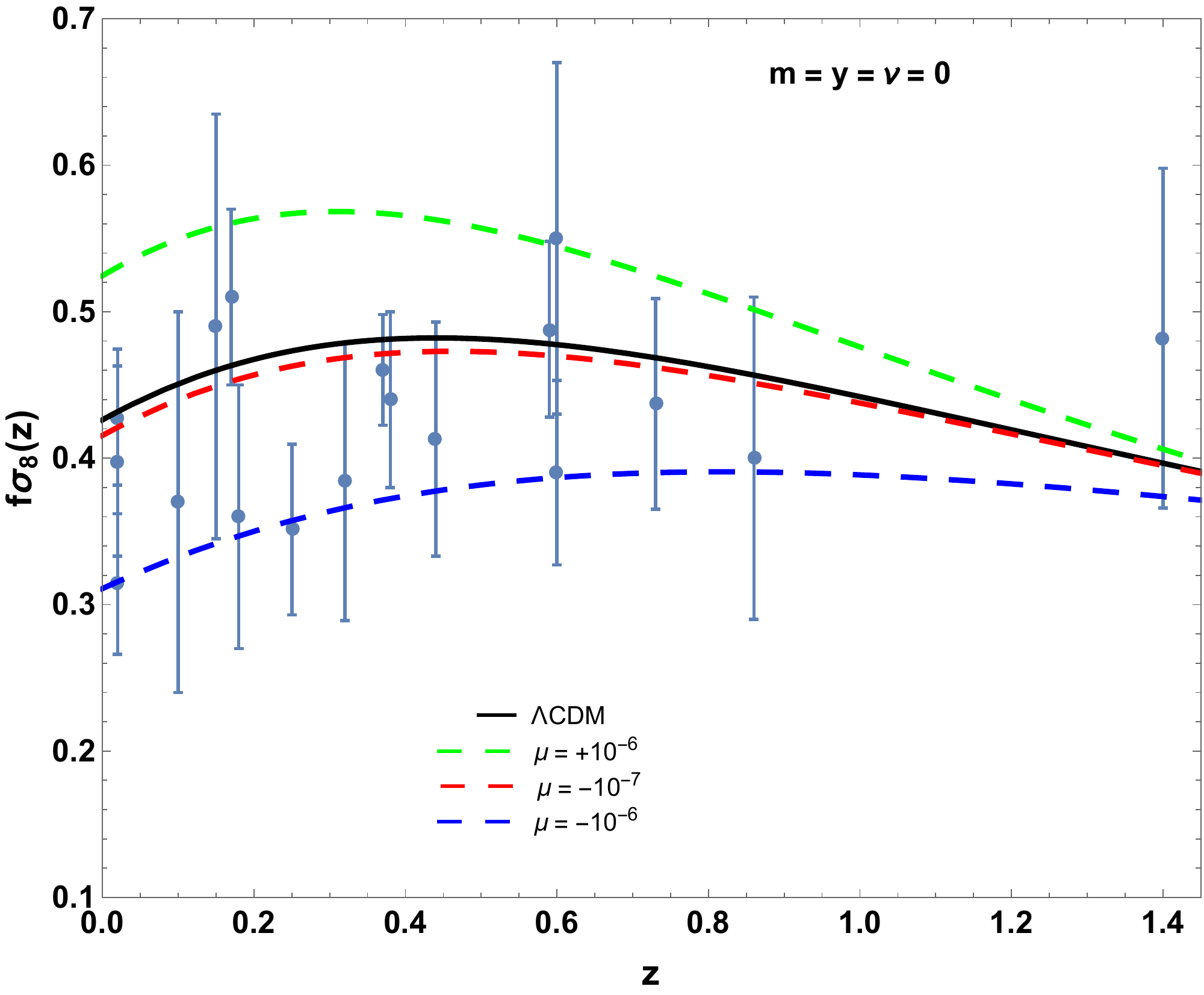}
\hspace{1cm}
\includegraphics[width=0.45\textwidth]{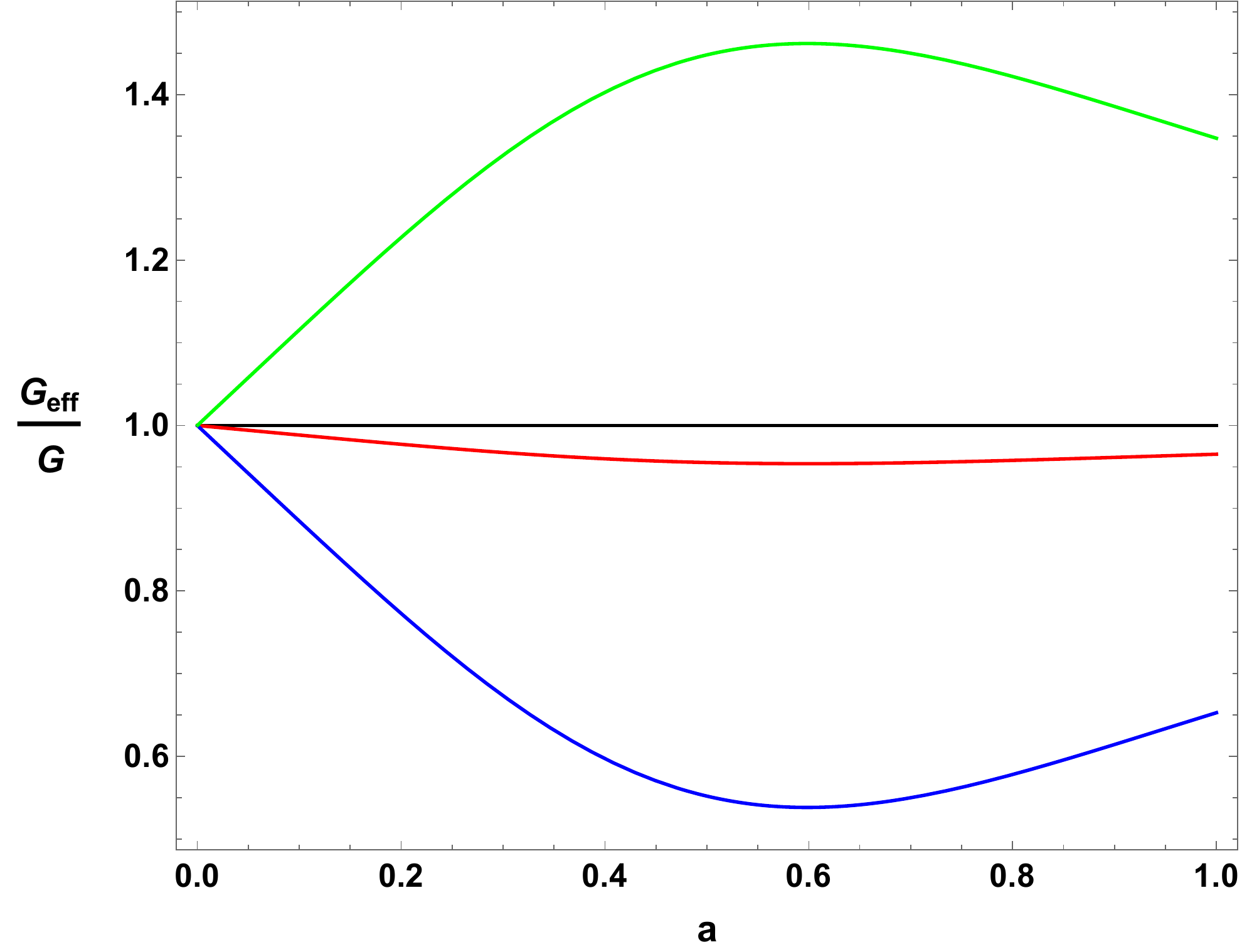}
\label{}	
\caption{Dependence of $f\sigma_8 (z)$ on $z$ for the $\Lambda$CDM background value $m=0$ with $\nu = y=0$. The different curves correspond to different values of the anisotropy parameter $\mu$. Negative values of $\mu$ lower the theoretical curve compared with the $\Lambda$CDM result, positive values of $\mu$ heighten the curve.
The right panel shows the scale-factor dependence of the effective gravitational ``constant". }
\end{center}
\end{figure}

\begin{figure}[h!]
\begin{center}
\includegraphics[width=0.45\textwidth]{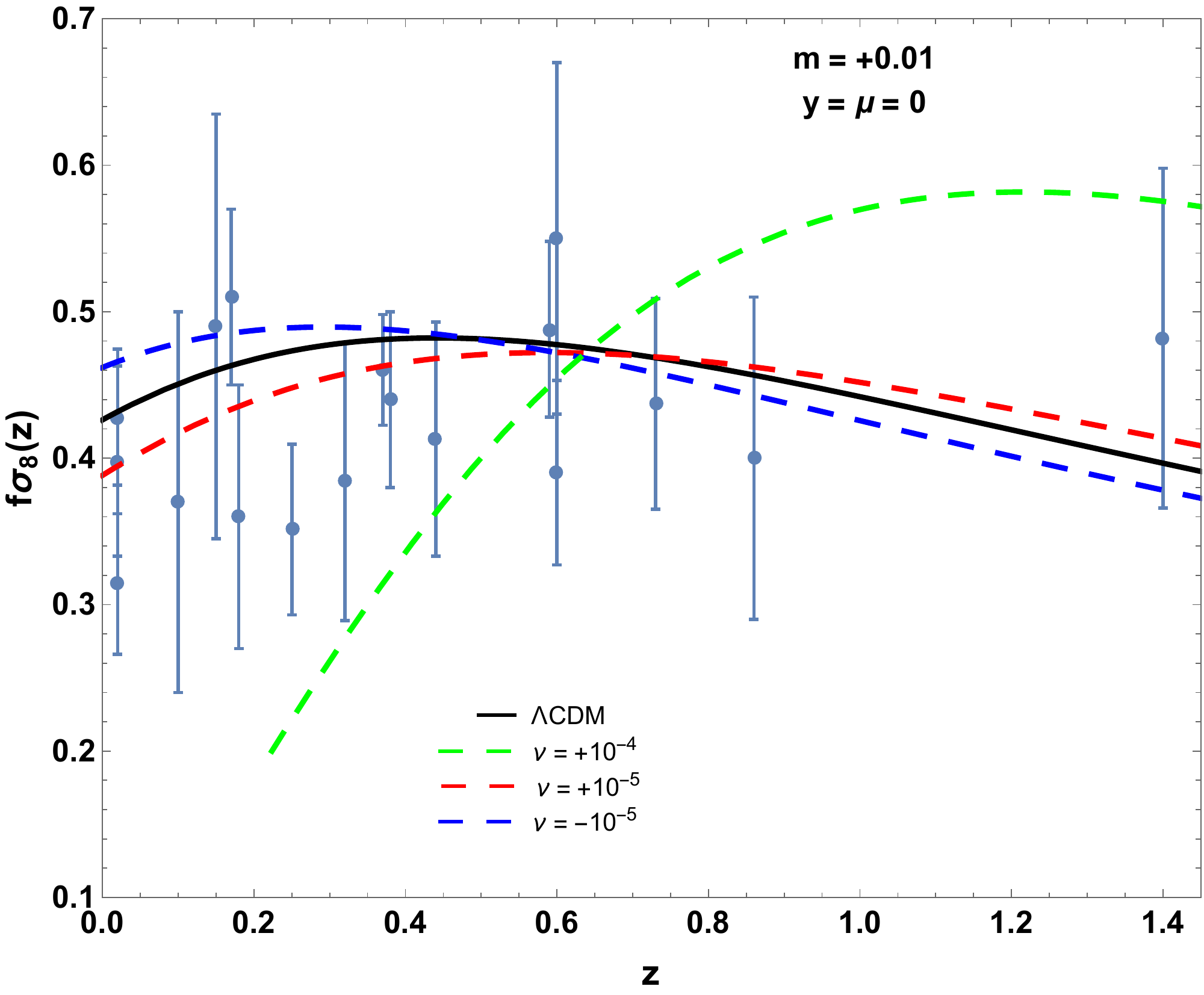}
\hspace{1cm}
\includegraphics[width=0.45\textwidth]{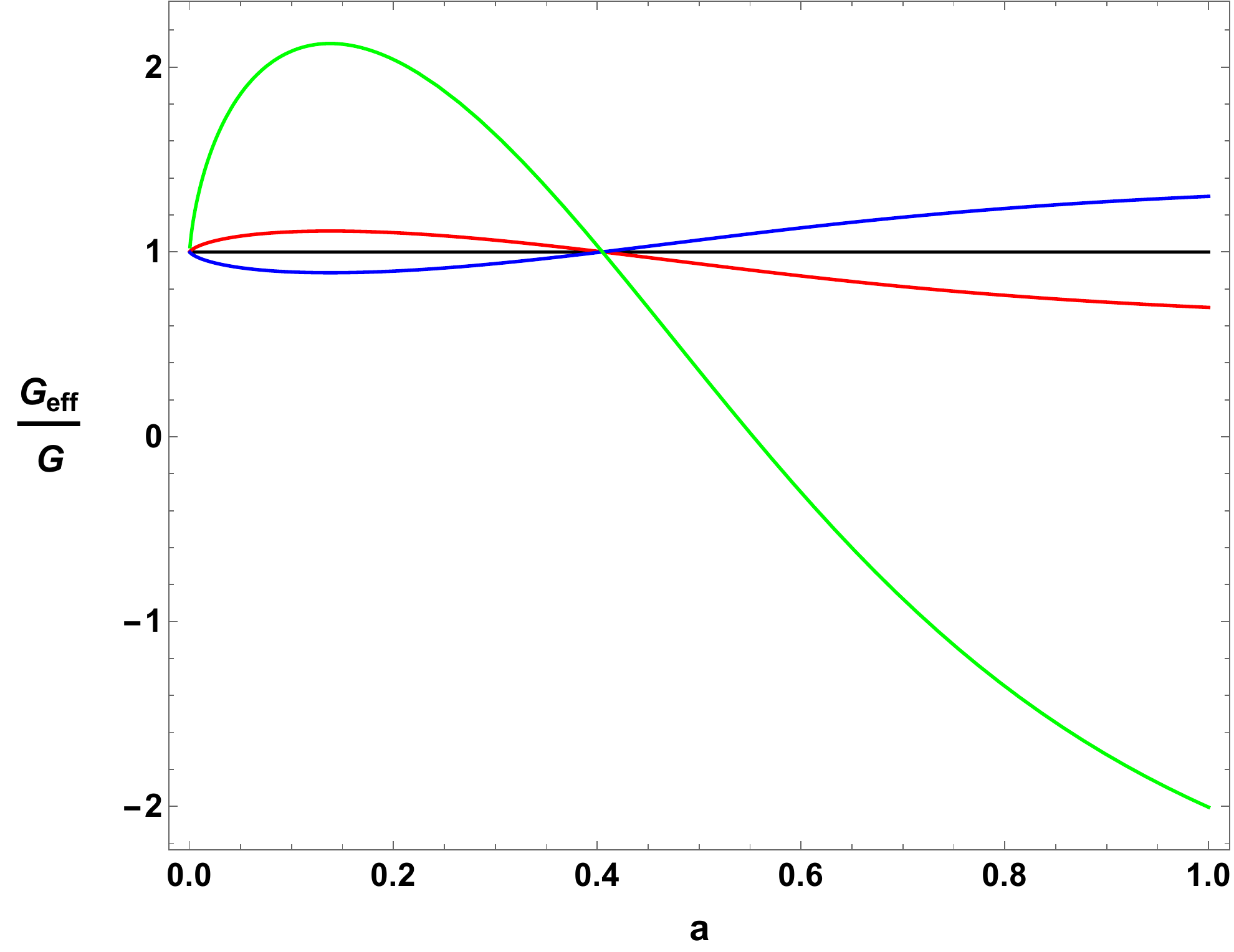}
\label{}	
\caption{Dependence of $f\sigma_8 (z)$ on $z$ for the fixed non-standard background value $m=0.01$ with $\mu= y=0$. The different curves correspond to different values of the anisotropy parameter $\nu$. Up to values of the order of $\nu \approx \pm 10^{-5}$ the deviation from the standard model remains small. For $\nu = 10^{-4}$ it becomes unacceptably large. The effective gravitational ``constant" (right panel) may change between values larger and smaller than the standard value during the cosmic expansion. }
\end{center}
\end{figure}

\begin{figure}[h!]
\begin{center}
\includegraphics[width=0.45\textwidth]{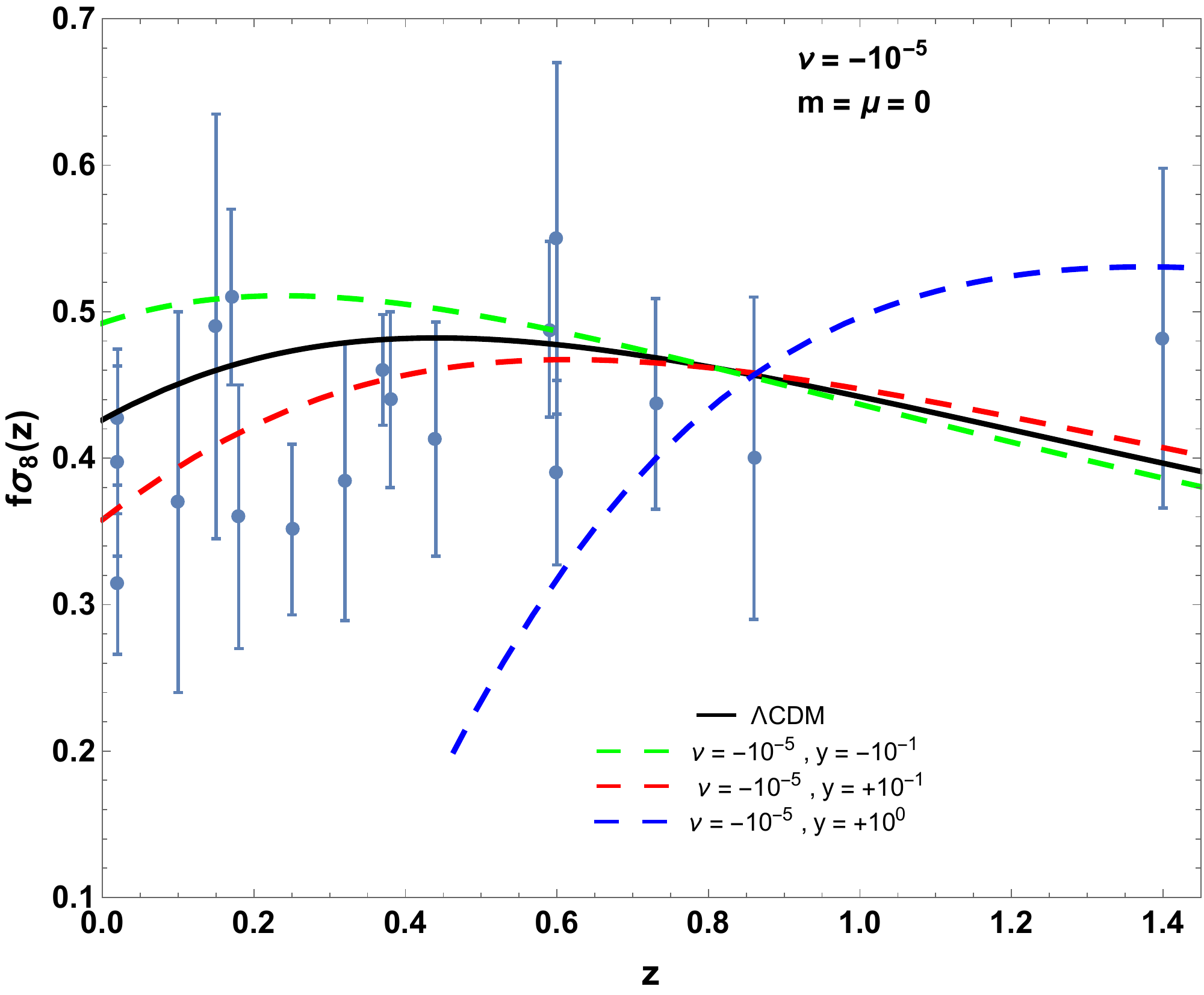}
\hspace{1cm}
\includegraphics[width=0.45\textwidth]{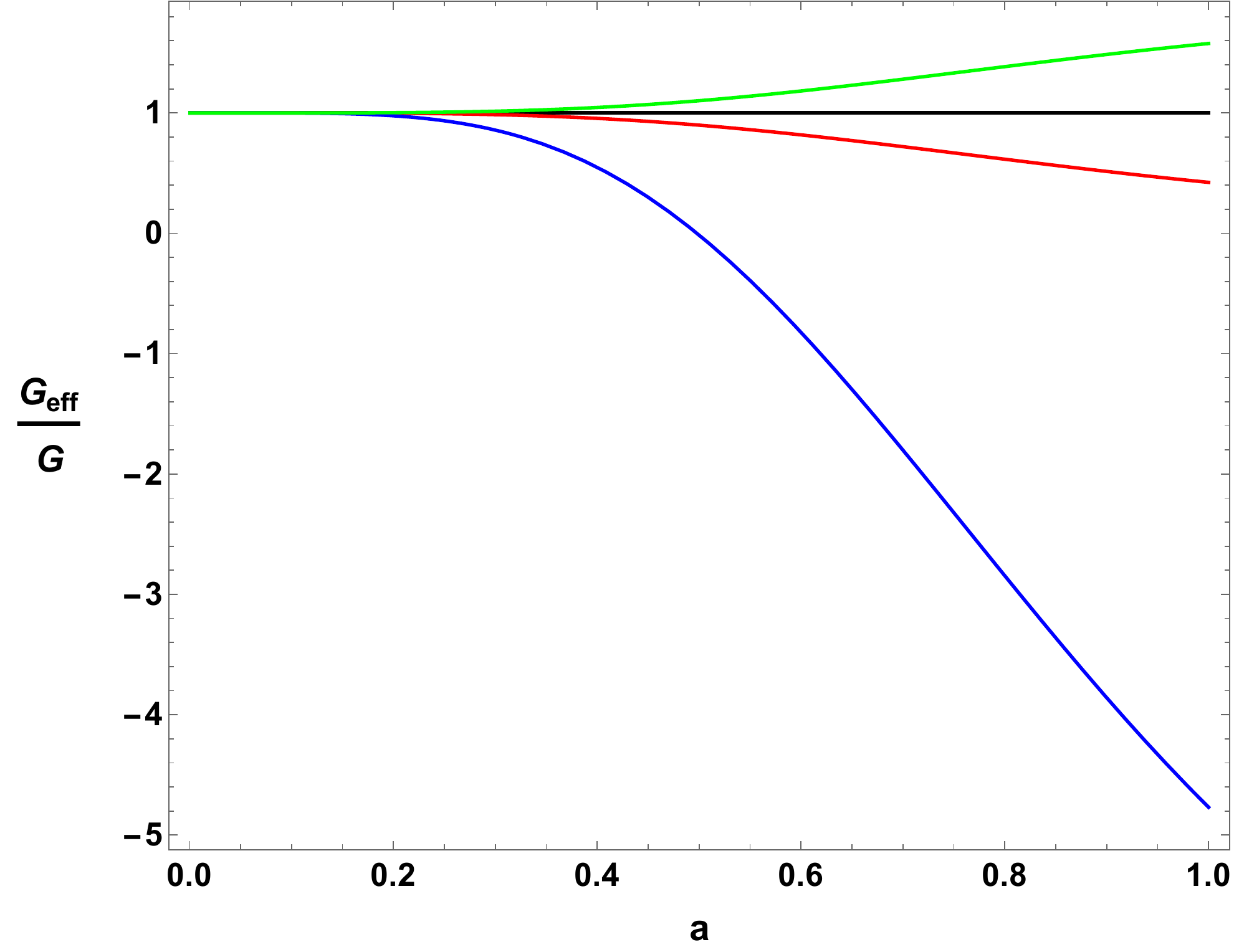}
\label{}	
\caption{Dependence of $f\sigma_8 (z)$ on $z$ for the standard-model background value $m=0$ with $\nu=-10^{-5}$ and $\mu=0$. The different curves correspond to different contributions from DE perturbations. This figure describes a cross effect between anisotropic stress, induced by relative perturbations, and DE perturbations. Here, positive values of $y$ lower (for $z\lesssim 1$) the curve for $f\sigma_8$, corresponding to a reduced effective gravitational ``constant" (right panel). }
\end{center}
\end{figure}

%%%%%%%%%%%%%%%%%%%%%%%%%%%%%%%%%%%%%%%%%%%%%%%%%%%%%%%%%%%%%%%%%%%%%%%%%%%%%%%%%%%%%%%%%%%%%%%%%%%%%%%%%%%%%%%%%%%%%%%%%%%%%%%%%%%%%%%%%%%%%%%%%%%%%%%%%%%%%%%%%%%%%%%%%%%%%%%%

\section{Discussion}
\label{discussion}

Adopting an effective GR framework and using a simple parametrization of geometric DE perturbations we have established a fluid-dynamical description of the matter growth in a JBD inspired extension of the $\Lambda$CDM model.
In a homogeneous and isotropic background this extension, which we call $e_{\Phi}\Lambda$CDM model, is characterized by an explicit analytic expression for the effective Hubble rate in which deviations from the standard model  are  taken into account by the additional parameter $m$, where $m=0$ represents the $\Lambda$CDM limit.
On the basis of this extension we studied the separate and the combined influences  of deviations from the $\Lambda$CDM background, of anisotropic stresses and of non-vanishing DE perturbations.
Combining the fluid conservation equations with the Raychaudhuri equation for the expansion scalar, the complete  first-order, scalar perturbation dynamics can be condensed into a manifestly gauge-invariant coupled system of two second-order differential equations for the total and the relative energy-density perturbations.
This background-independent perturbation dynamics is formulated entirely in terms of (gauge-invariantly defined) energy densities and pressures. Neither velocity components nor  metric functions do appear explicitly.
To close the system, assumptions about the isotropic and anisotropic pressure perturbations have to be made. For the scalar part of the anisotropic pressure (which measures the gravitational slip) we postulate a proportionality to the total and to the relative
energy-density perturbations, extending the usual procedure for the isotropic pressure in the
simplest possible way. From the solutions of the system the observationally relevant matter growth rate is obtained straightforwardly.
Deviations of the background dynamics alone lead to small corrections of the $\Lambda$CDM prediction for $f\sigma_8$ which are  largest at a redshift of about $0.5$.
Our analysis suggests that at the chosen scale $k= 0.1 h \mathrm{Mpc}^{-1}$ DE perturbations have to be at least one order of magnitude smaller than matter perturbations.
Negative values of the anisotropy parameter $\mu$ which quantifies the coupling of the anisotropic stress to the total density fluctuations lower the theoretical curve for  $f\sigma_8$ compared with the $\Lambda$CDM result. This corresponds to a reduction of the effective gravitational constant.  Positive values of  $\mu$ act in the opposite direction and seem to be less favored by the data.
The coupling to  relative perturbations is described by the parameter $\nu$. Here, negative values heighten the theoretical curve at small redshift, while positive values lower it. During the cosmic evolution the effective gravitational ``constant" may change between values larger and smaller than the standard value.
Values of the order of $|\nu|\lesssim 10^{-5}$ are compatible with current data while $\mu$ is restricted to $|\mu|\lesssim 10^{-6}$. In other words, for a $\nu$ of the same order as $\mu$ the anisotropic stresses related to the total density perturbations (parametrized by $\mu$) have a larger impact on the effective gravitational ``constant" than the anisotropic stresses generated by the relative perturbations (parametrized by $\nu$).
There exists a cross effect between anisotropic stresses, induced by relative perturbations, and DE perturbations.
Here, perturbations $y>0$ tend to lower the curve for $f\sigma_8$ at small redshift which is opposite to the impact of DE perturbations in the absence of anisotropic pressures.
Obviously, in the preliminary analysis of this paper there is some degeneracy at low redshift in the parameter space spanned by $\mu$, $\nu$ and $y$. A detailed analysis of the complete system  of coupled perturbation equations for $\delta^{c}$ and $S_{m}$ will be the subject of a subsequent paper.
\ \\
\ \\
\noindent
{\bf Acknowledgement:} Financial support by  CNPq, CAPES and FAPES is gratefully acknowledged.
%%%%%%%%%%%%%%%%%%%%%%%%%%%%%%%%%%%%%%%%%%%%%%%%%%%%%%%%%%%%


\begin{thebibliography}{99}

\bibitem{WHW} W.C. Algoner, H.E.S. Velten and W. Zimdahl,
\textit{Scalar-tensor extension of the $\Lambda$CDM model}, JCAP \textbf{1611}, 034 (2016).

%%%%%%%%%%%%%%%%%%%%%%%%%%%%%%%%%%%%%%%%%%%%%%%%%%%%%%%%%%%%%%%%%%%%%%%%%%%%%%%%%%%%%%%%%%%%%%%%%%%%%%%%%%%%%%%%%%%%%%%%%%%%%%%%
\bibitem{jordan} P. Jordan, Z. Physik \textbf{157}, 112 (1959).

\bibitem{BD} C. Brans and R.H. Dicke, Phys.Rev. \textbf{124}, 925 (1961).

\bibitem{B} R.H. Dicke, Phys.Rev. \textbf{125}, 2163 (1962).


 \bibitem{nagata} R. Nagata, T. Chiba and N. Sugiyama,
\textit{Observational consequences of the evolution of primordial fluctuations in scalar-tensor cosmology},
Phys.Rev.D \textbf{66}, 103510 (2002).

\bibitem{catena} R. Catena, M. Pietroni and L. Scarabello, {\it Einstein and Jordan reconciled: a frame-invariant approach to scalar-tensor cosmology}, Phys.Rev.D \textbf{76}, (2007) 084039,
arXiv:astro-ph/0604492.

\bibitem{farao} V. Faraoni and S. Nadeau, {\it The (pseudo)issue of the conformal frame revisited}, Phys.Rev.D \textbf{75}, (2007) 023501,
arXiv:gr-qc/0612075.


\bibitem{dolgov} A. D. Dolgov and M. Kawasaki, {\it  	
Can modified gravity explain accelerated cosmic expansion?}, Phys.Lett. B {\bf 573},
1 (2003).

\bibitem{chiba} T. Chiba, {\it  	
1/R gravity and scalar - tensor gravity}, Phys.Lett.B {\bf 575},
1 (2003).

\bibitem{duma} R. Durrer and R. Maartens, {\it Dark Energy: Observational \& Theoretical Approaches}, ed. P Ruiz-Lapuente (Cambridge UP, 2010), 48,
arXiv: 0811.4132.

\bibitem{sofarao} T.P. Sotiriou and V. Faraoni, {\it  	
f(R) Theories Of Gravity}, Rev.Mod.Phys. 82 (2010) 451-497,
arXiv:0805.1726.


\bibitem{abean} N. Agarwal and R. Bean, {\it  	
The Dynamical viability of scalar-tensor gravity theories},  Class.Quant.Grav.\textbf{25}, (2008) 165001, arXiv:0708.3967.

\bibitem{scalten} C.E.M. Batista and W. Zimdahl, {\it Power-law solutions and accelerated expansion in scalar-tensor theories}, Phys.Rev.D \textbf{82}, 023527 (2010); arXiv:0912.0998.



\bibitem{clifton} T. Clifton, P. Ferreira, A. Padilla and C. Skordis, 	
\textit{Modified Gravity and Cosmology}, Physics Reports \textbf{513} (2012) 1-189.




\bibitem{HwangNoh2005} Jai-chan Hwang and Hyerim Noh, \textit{Classical evolution and quantum generation in generalized gravity theories including string
corrections and tachyons: Unified analyses}, Phys.Rev.D \textbf{71}, 063536 (2005).

\bibitem{battye} R.A. Battye and J.A. Pearson, \textit{Effective action approach to
cosmological perturbations in dark energy and modified gravity}, JCAP 1207, 019 (2012).

\bibitem{battye13} R.A. Battye and J.A. Pearson, \textit{Parametrizing dark sector perturbations via equations of state}, Phys.Rev.D \textbf{88}, 061301(R) (2013).

\bibitem{battye16} R.A. Battye, B. Bolliet and J.A. Pearson, \textit{f($\cal{R}$) gravity as a dark energy fluid},
 Phys.Rev.D \textbf{93}, 044026 (2013).

\bibitem{chibayam} T. Chiba and M. Yamaguchi, {\it Conformal-Frame (In)dependence of Cosmological Observations in Scalar-Tensor Theory} JCAP 1310, (2013) 040.

\bibitem{joyce} A. Joyce, L. Lombriser and F. Schmidt, {\it Dark Energy vs. Modified Gravity}, Ann.Rev.Nuc.Part.Sc. 2016. AA:1-28; arXiv:1601.06133.


\bibitem{torres} D. Torres, \textit{Quintessence, superquintessence and observable quantities in Brans-Dicke and
nonminimally coupled theories},
Phys.Rev.D \textbf{66}, 043522 (2002), arXiv:astro-ph/0204504.

\bibitem{carroll}
S. M. Carroll, V. Duvvuri, M. Trodden and M.S. Turner, {\it  	
Is cosmic speed - up due to new gravitational physics?} Phys. Rev. D \textbf{70}, 043528 (2004), arXiv:astro-ph/0306438.

\bibitem{nojiri1} S. Nojiri and S.D. Odintsov, {\it Where new gravitational physics comes from: M-theory?} Phys. Lett. B \textbf{576}, 5 (2003), arXiv:hep-th/0307071; {\it Introduction to Modified Gravity and Gravitational Alternative for Dark Energy},
Int.J.Geom.Meth.Mod.Phys. \textbf{4}, 115 (2007), arXiv:hep-th/0601213.

\bibitem{gannouji1} R. Gannouji, D. Polarski, A. Ranquet and A.A. Starobinsky, {\it Scalar-Tensor Models of Normal and Phantom Dark Energy},
JCAP \textbf{0609}, 016 (2006), arXiv:astro-ph/0606287.

\bibitem{copeland}
E.J. Copeland, M.Sami and S. Tsujikawa, {\it Dynamics of dark energy}, Int.J.Mod.Phys.D \textbf{15}, 1753 (2006), arXiv:hep-th/0603057.



\bibitem{lobo} F.S.N. Lobo, {\it The dark side of gravity: Modified theories of gravity}, Research Signpost, ISBN 978-81-308-0341-8,  173 (2009), arXiv:0807.1640.

\bibitem{caldkam} R.C. Caldwell and M. Kamionkowski, {\it  	
The Physics of Cosmic Acceleration}, Ann.Rev.Nucl.Part.Sci.{\bf 59}, 397 (2009), arXiv:0903.0866.

\bibitem{baker} T. Baker, P.G. Ferreira and
C. Skordis,\textit{ The parameterized post-Friedmann framework for theories of modified gravity:
Concepts, formalism, and examples},  Phys.Rev.D  \textbf{87}, 024015 (2013).

\bibitem{limaferreira} N.A. Lima and P.G. Ferreira, \textit{On the phenomenology of extended
Brans-Dicke gravity},  JCAP \textbf{1601}, 010 (2016).

\bibitem{kofinaslima} G. Kofinas and N.A. Lima, \textit{Dynamics of cosmological perturbations in modified Brans-Dicke cosmology with
matter-scalar field interaction}, arXiv:1704.08925.


\bibitem{wuchen} Feng-Quan Wu and Xuelei Chen, \textit{Cosmic microwave background with Brans-Dicke gravity. II. Constraints
with the WMAP and SDSS data},

\bibitem{liwuchen} Yi-Chao Li, Feng-Quan Wu and Xuelei Chen, \textit{Constraints on the Brans-Dicke gravity theory with the Planck data}, Phys.Rev.D  \textbf{88}, 084053 (2013).

\bibitem{hrycyna} O. Hrycyna, M. Szyd{\l}owski and M. Kamionka, \textit{Dynamics and cosmological constraints on Brans-Dicke cosmology}, Phys.Rev.D  \textbf{90}, 124040 (2014).

\bibitem{avilezskordis} A. Avilez and C. Skordis, \textit{Cosmological Constraints on Brans-Dicke Theory}, PRL \textbf{113}, 011101 (2014).

\bibitem{umilta} C. Umilt\`{a},  M. Ballardini, F. Finelli  and D. Paoletti, \textit{CMB and BAO constraints for an
induced gravity dark energy model with a quartic potential}, JCAP \textbf{1508}, 017 (2015).

\bibitem{alonsobellini} D. Alonso, E. Bellini, G. Ferreira, and M. Zumalac\'{a}rregui, \textit{Observational future of cosmological scalar-tensor theories},
Phys.Rev.D  \textbf{95}, 063502 (2017).

\bibitem{LAmend} I. Sawicki, I.D. Saltas, L. Amendola and M. Kunz,
\textit{Consistent perturbations in an imperfect fluid}, JCAP \textbf{1301}, 004 (2013).

\bibitem{nesseris17} S. Nesseris, G. Pantazis, L. Perivolaropoulos, \textit{Tension and constraints on modified gravity parametrizations of Geff(z) from growth rate and Planck data},
arXiv:1703.10538.

\bibitem{baVDF} W.S. Hip\'olito-Ricaldi, H.E.S. Velten and W. Zimdahl, 2010, Phys.Rev.D  \textbf{82}, 063507 (2010).
.
\bibitem{ricci} S. del Campo, J.C. Fabris, R. Herrera and W. Zimdahl,
\textit{Cosmology with Ricci dark energy},
Phys.Rev.D \textbf{87}, 123002 (2013).

\bibitem{alonso}
A. Romero Fu\~{n}o, W.S. Hip\'{o}lito-Ricaldi, W. Zimdahl, {\it Matter perturbations in scaling cosmology}.
MNRAS \textbf{457}, 2958 (2016); arXiv:1409.7706.

\bibitem{kunzsapone07} M. Kunz and D. Sapone, \textit{Dark Energy versus Modified Gravity},
Phys.Rev.Lett. \textbf{98}, 121301 (2007).

\bibitem{cardona} W. Cardona,,L. Hollenstein and M. Kunz,
\textit{The traces of anisotropic dark energy in light of Planck},
JCAP \textbf{1407}, 032 (2014).

\bibitem{blas} D. Blas, S. Floerchinger, M. Garny,
N. Tetradis and U.A. Wiedemann, \textit{Large scale structure from viscous dark matter},
JCAP \textbf{1511}, 049 (2015).

\bibitem{faraoni} V. Faraoni and J. Cot\'{e}, Imperfect fluid description of modified gravities, arXiv:1808.02427.

\bibitem{boisseau}
B. Boisseau, G. Esposito-Farese, D. Polarski, and A. A.
Starobinsky, \textit{Reconstruction of a Scalar-Tensor Theory of Gravity in an Accelerating Universe},
Phys.Rev.Lett. \textbf{85}, 2236 (2000).

\bibitem{tsujikawa07} S. Tsujikawa,
\textit{Matter density perturbations and effective gravitational constant in modified gravity models
of dark energy},
Phys.Rev.D \textbf{76}, 023514 (2007).

\bibitem{bertschinger} E. Bertschinger and Ph. Zukin, \textit{Distinguishing modified gravity from dark energy},
Phys.Rev.D \textbf{78}, 024015 (2008).

\bibitem{gannouji} R. Gannouji and D. Polarski, \textit{The growth of matter perturbations in
some scalar–tensor DE models},
JCAP \textbf{0805}, 018 (2008).

\bibitem{song09} Yong-Seon Song and W.J. Percival, \textit{Reconstructing the history of structure
formation using redshift distortions},
JCAP \textbf{0910}, 009 (2009).


\bibitem{bean} R. Bean and M. Tangmatitham, \textit{Current constraints on the cosmic growth history},
Phys.Rev.D  \textbf{81}, 083534 (2010).

\bibitem{hojjati} A. Hojjati, L. Pogosian and Gong-Bo Zhao, \textit{Testing gravity with CAMB and
CosmoMC}, JCAP \textbf{1108}, 005 (2011).

\bibitem{Silvestri} A. Silvestri, L. Pogosian and R.V. Buniy,
\textit{A practical approach to cosmological perturbations in modified gravity},
	arXiv:1302.1193;
Phys.Rev.D \textbf{87}, 104015 (2013).



\bibitem{steigerwald14} H. Steigerwald, J. Belb and Ch. Marinoni,
\textit{Probing non-standard gravity with the growth index: a background independent analysis},
JCAP \textbf{1405}, 042 (2014).

\bibitem{piazza14} F. Piazza, H. Steigerwald and Ch. Marinoni,
\textit{Phenomenology of dark energy: exploring the space of theories with future redshift surveys},
JCAP \textbf{1405}, 043 (2014).


\bibitem{pertsaulo}  W. Zimdahl, H.A. Borges, S. Carneiro, J.C. Fabris and W.S. Hip\'{o}lito-Ricaldi, {\it Non-adiabatic perturbations in decaying vacuum cosmology}.
JCAP \textbf{1104}, 028 (2011).



\bibitem{BuenoS} J. C. Bueno Sanchez and L. Perivolaropoulos, \textit{Evolution of dark energy perturbations in scalar-tensor cosmologies},
Phys.Rev.D  \textbf{81}, 103505 (2010).

\bibitem{SaltasKunz} I.D. Saltas and M. Kunz, \textit{Anisotropic stress and stability in modified gravity models},
Phys.Rev.D \textbf{83}, 064042 (2011).





%Anisotropic stress
\bibitem{DossettIshak} J.N. Dossett and M. Ishak, \textit{Effects of dark energy perturbations on cosmological tests of general relativity},
Phys.Rev.D \textbf{88}, 103008 (2013).

\bibitem{amendola13} L. Amendola, S. Fogli, A. Guarnizo, M. Kunz and A. Vollmer,
\textit{Model-independent constraints on the cosmological anisotropic stress}, Phys.Rev. D \textbf{89}, 063538 (2014);
arxiv:1311.4765



%Neuere Literatur zu ST





%%%%%%%%%%%%%%%%%%%%%%%%%%%%%%%%%%%%%%%%%%%%%%%%%%%%%%%%%%%%%%%%%%%%%%%%%%%%%%%%%%%%%%%%%%%%%%%%%%%%%%%%%%%%%%%%%%%%%%%%%%%%%%%%





\bibitem{silveirawaga} V. Silveira and I. Waga, \textit{Decaying $\Lambda$ cosmologies and power spectrum},
Phys.Rev.D 50, 4890 (1994).






\bibitem{abramo07} L.R. Abramo, R.C. Batista, L. Liberato and R. Rosenfeld,
\textit{Structure formation in the presence of
dark energy perturbations},
JCAP \textbf{0711}, 012 (2007).

\bibitem{saponekunz09} D. Sapone and M. Kunz, \textit{Fingerprinting dark energy},
Phys.Rev.D 80, 083519 (2009).

\bibitem{Park-Hwang} C.-G. Park, J. Hwang, J. Lee, and H. Noh,
Phys.Rev.Lett. \textbf{103}, 151303 (2009).

\bibitem{song} Yong-Seon Song, L. Hollenstein, G. Caldera-Cabral and K. Koyama,
\textit{Theoretical priors on modified growth parametrisations},
JCAP 1004 018 (2010); arXiv:1001.0969.

\bibitem{nesseris15} S Nesseris, and D. Sapone, \textit{Accuracy of the growth index in the presence of dark energy perturbations},
Phys.Rev.D 92, 023013 (2015).

\bibitem{basilakos15} S. Basilakos, \textit{The growth index of matter perturbations using the clustering
of dark energy},
MNRAS \textbf{449}, 2151–2155 (2015).

\bibitem{nesseris08} S. Nesseris and L. Perivolaropoulos
\textit{Testing LCDM with the Growth Function $\delta(a)$: Current Constraints},
Phys.Rev.D \textbf{77}, 023504 (2008); arXiv:0710.1092.

\bibitem{Linder16} E.V. Linder, \textit{Cosmic Growth and Expansion Conjoined},
Astroparticle Physics \textbf{86}, 41 (2017); arXiv:1610.05321.
%%%%%%%%%%%%%%%%%%%%%%%%%%%%%%%%%%%%%





\end{thebibliography}
\end{document}